\documentclass[journal]{IEEEtran}
\usepackage[utf8]{inputenc}
\usepackage[T1]{fontenc}
\usepackage{cite,amsmath,amssymb,graphicx,booktabs,array,longtable,url,placeins}
\usepackage[hidelinks]{hyperref}
\makeatletter
\newcommand\cleanbarrier{\par\ifx\@deferlist\@empty\ifx\@dbldeferlist\@empty\else\@cbclear\fi\else\@cbclear\fi}
\newcommand\@cbclear{\clearpage\global\@colroom\@colht\global\vsize\@colht}
\makeatother
\begin{document}
\title{Smart Charging Without Rebound Peaks: Wholesale Price Feedback Limits Distribution Grid Peaks From Price-Responsive Electric Vehicles}
\author{Lion~Hirth
\thanks{L. Hirth is with the Hertie School, Berlin, Germany, and Neon Neue Energieökonomik GmbH, Berlin, Germany (e-mail: hirth@hertie-school.org).}%
\thanks{This version: 30 October 2026. JEL classification: Q41, Q48, L94, D47, C63.}%
\thanks{This work has been submitted to the IEEE for possible publication. Copyright may be transferred without notice, after which this version may no longer be accessible.}}
\maketitle
\begin{abstract}
Studies on price-optimized EV charging consistently find that the common price signal synchronizes charging of the entire fleet of cars in the cheapest hours and creates new ``rebound'' load peaks in low-voltage grids. Almost all of these studies treat wholesale prices as exogenous, i.e., unaffected by the shifting EV load. This paper argues that this assumption overstates the problem: charging in low-price hours raises the wholesale price in those hours, which spreads charging across hours in equilibrium. I simulate price-optimized charging under real-time prices calibrated to German 2025 data. Even when adjusting for the increase in electricity demand, the coincident import peak of the local grid rises by 260\,\%. After adding a model of equilibrium wholesale prices that takes the price-quantity feedback into account, the increase is reduced to a mere 5\,\%. Allowing for vehicle-to-grid, the export peak rises by 115\,\% under exogenous prices and \emph{falls} by 6\,\% under endogenous prices. The result holds across vehicle and charging parameters, charging constraints, renewable scenarios, network charges, price functions, peak measures and simulation assumptions. Rebound peaks remain possible where vehicle adoption runs ahead of the national average or where charging power is high, but they are much smaller than the literature consensus suggests.
\end{abstract}
\section{Introduction}\label{sec:1}
Electric vehicles are the largest flexible load that households connect to the low-voltage grid. A battery electric car stores 40--100\,kWh and charges at 11\,kW on a three-phase home connection, more than the peak load of a typical single-family home. Most charging takes place at home, often overnight, and the energy for a typical day of driving is recharged in less than an hour. Where vehicle owners pay a retail tariff that follows the hourly wholesale price (real-time pricing), they have an incentive to charge when wholesale prices are low. From a system perspective this is efficient: low prices signal abundant low-cost generation and/or low demand.

Distribution grid operators are concerned about that very mechanism, because every vehicle in a grid responds to the same national price signal. If all vehicles charge at full power in the same cheapest hour of the night, the aggregate load of a feeder or transformer can exceed the evening peak it was built for. The literature calls this ``herding'' or a ``rebound peak'' \cite{r1,r2,r3,r4,r5,r6,r7}. Almost every simulation study of price-responsive charging finds a rebound, often several times larger than in no-EV or uncontrolled-charging benchmarks. In response, the literature proposes staggered time-of-use windows \cite{r4,r6,r8}, capacity-based or dynamic network charges \cite{r1,r9,r10,r11,r12}, capacity limits \cite{r7,r13} or direct control of charging \cite{r14,r15,r16,r17,r18,r19}.

Most of these simulations share one crucial assumption: the prices that EVs face are exogenous, i.e., the price series is fixed before the vehicles are dispatched, and the dispatch does not change it. In the terminology of the control literature, prices are open-loop \cite{r20,r21,r22,r23}. For a single vehicle this is reasonable. For a national fleet of millions of vehicles it is not: charging in the cheapest hours adds demand in exactly those hours, which raises the wholesale price and dampens the incentive to charge then. Prices and charging are determined jointly in a market equilibrium, and the price feedback spreads charging across more hours. Of the 38 simulation studies reviewed in this paper, 34 treat prices as exogenous. This paper asks whether the rebound peak persists under endogenous prices and how much the price feedback reduces it.

The paper makes four contributions. First, based on a broad literature search, it identifies and reviews 44 studies on price-responsive vehicle charging and local grid stress, assessing for each whether prices were modeled as endogenous or exogenous. As a side benefit, it provides a comprehensive overview of quantifications of the rebound peak. Second, it applies a simple closed-loop model of wholesale prices to electric vehicles: the wholesale price is a function of national residual load including vehicle charging and discharging, and the equilibrium is computed as a fixed point. Third, it presents simulation results for 100 households on one low-voltage grid, calibrated to German data for 2025, with vehicle-to-grid, mobility constraints and a growing share of price-optimized vehicles. Fourth, to assess the robustness of results, it varies vehicle and charging parameters, charging constraints, the local concentration of vehicles, renewable supply, network charges, price functions and peak measures.

The main finding is that the price feedback removes the rebound peak almost entirely. With exogenous prices, the simulations reproduce the literature: once more than 5--10\,\% of households charge a vehicle against the price, the import peak of the local grid exceeds its no-EV level, even after adjusting for the increase in overall electricity consumption. At 50\,\% it is 260\,\% higher. At the same penetration, with endogenous prices, the same fleet raises the import peak by just 5\,\%. In many real-world low-voltage grids, a 260\,\% peak increase would imply large grid expansion costs and a replacement of the transformer, while 5\,\% is usually well within capacity margins.

In the export direction, vehicle-to-grid discharging raises the peak by 115\,\% under exogenous prices. It \emph{lowers} it by 6\,\% under endogenous prices. This suggests V2G helps absorb local solar peaks, while exogenous-price studies suggest the opposite.

The (national) wholesale price response is an effective mechanism against local grid peaks if EV adoption is uniform across the country. If EV adoption is locally concentrated, e.g., in affluent neighborhoods or environmentally conscious communities, rebound peaks emerge, the simulations suggest. It is in this context that grid instruments, such as grid operator curtailment or congestion-based network charges, might be necessary.

The next section reviews the literature. Section~\ref{sec:3} explains the price feedback, Section~\ref{sec:4} the methodology. Section~\ref{sec:5} presents the results, Section~\ref{sec:6} robustness checks, and Section~\ref{sec:7} concludes.

\section{Related work}\label{sec:2}
From an initial pool of more than 300 papers I identified 44 studies that quantify the impact of price-responsive electric vehicles on local distribution grids. Of these, 38 are simulations and six are empirical studies based on field experiments or smart meter data; they are listed in Tables~\ref{tab:1} and~\ref{tab:2} (details in Tables~\ref{tab:A1} and~\ref{tab:A2} in the Appendix). Twenty-seven studies use real-time pricing, eleven time-of-use tariffs and six other price-based signals such as imbalance prices or local prices. Studies use different grid impact metrics, mostly the peak load of a feeder or transformer, but also overload hours or probabilities, voltage, or reinforcement cost. For the purpose of this paper, all of them indicate stress on the local low-voltage grid.

\begin{table*}[!tp]
\caption{Simulation studies on price-responsive electric vehicle charging and local grid stress (details in Table~\ref{tab:A1})}\label{tab:1}
\centering\footnotesize\setlength{\tabcolsep}{3pt}
\begin{tabular}{@{}>{\raggedright\arraybackslash}p{0.216\linewidth}>{\raggedright\arraybackslash}p{0.101\linewidth}>{\raggedright\arraybackslash}p{0.086\linewidth}>{\raggedright\arraybackslash}p{0.316\linewidth}>{\raggedright\arraybackslash}p{0.123\linewidth}>{\raggedright\arraybackslash}p{0.078\linewidth}@{}}
\toprule
\textbf{Study} & \textbf{Country} & \textbf{Signal} & \textbf{Grid metric} & \textbf{New peak?} & \textbf{Price feedback} \\
\midrule
Stute and Kühnbach \cite{r24} & Germany & RTP, TOU & Household peak; transformer loading; overload hours & Partly & No \\
Semmelmann et al. \cite{r25} & Germany & RTP & Reinforcement cost & Yes & No \\
Cong et al. \cite{r26} & Denmark & RTP & Transformer loading; overload hours & Yes & No \\
Lilienkamp and Namockel \cite{r27} & Germany & RTP, TOU & Transformer overload events & Yes & No \\
Lersch et al. \cite{r28} & Germany & RTP & Transformer and line loading; voltage; reinforcement cost & Yes & No \\
Turk et al. \cite{r1} & US & TOU & Feeder peak & Yes & No \\
Verzijlbergh et al. \cite{r29} & Netherlands & RTP & Feeder peak & Yes & Yes \\
Stawska et al. \cite{r30} & Netherlands & Other & Feeder overload hours & Yes & No \\
Flath et al. \cite{r31} & Germany & Other & Distribution grid peak & Yes & No \\
Huang and Sandström \cite{r32} & Sweden & RTP & Transformer and cable loading; voltage & Yes & No \\
Salah et al. \cite{r33} & Switzerland & RTP & Substation loading & Yes & Yes \\
von Bonin et al. \cite{r34} & Germany & RTP & Aggregate peak, 100 households & Yes & No \\
Panda et al. \cite{r35} & Netherlands & RTP & Aggregate peak per charging point & Yes & No \\
Flath and Gottwalt \cite{r20} & Germany & Other & Aggregate peak, 900 households & Yes (open loop); no (closed loop) & Yes \\
Hennig et al. \cite{r36} & Germany & RTP & Transformer loading & Yes & No \\
Jones et al. \cite{r2} & US & TOU & Feeder peak & Yes & No \\
FTI Consulting \cite{r9} & EU & RTP & Feeder peak & Yes & No \\
Heinrich et al. \cite{r37} & Denmark & RTP & Primary substation loading & Yes & No \\
Ziras et al. \cite{r38} & Denmark & RTP & Aggregate load, 200 chargers & Yes & No \\
Unterluggauer et al. \cite{r39} & Denmark & RTP & Transformer and cable overload probability & Partly & No \\
Li et al. \cite{r40} & Denmark & RTP & Medium-voltage feeder loading & Yes & No \\
Hennig et al. \cite{r13} & Netherlands & RTP & Aggregate load, 50 households & Yes & No \\
Wu et al. \cite{r8} & Australia & TOU & Transformer utilization & Yes & No \\
Hein et al. \cite{r41} & Germany & RTP & Transformer and line loading; voltage & Yes & No \\
Lundblad et al. \cite{r42} & Sweden & RTP & Peak power; transformer and cable overloads; voltage & Yes & No \\
Jung and Sundström \cite{r43} & Sweden & RTP & Voltage variation & Yes & No \\
Wu et al. \cite{r44} & Australia & TOU & Transformer utilization & Yes & No \\
Shrestha et al. \cite{r45} & US & TOU & Feeder peak; transformer overloads; upgrade cost & Yes & No \\
Elmallah et al. \cite{r46} & US & TOU & Feeder and substation capacity; upgrade cost & Yes & No \\
Unterluggauer et al. \cite{r47} & Denmark & Other & Aggregate peak; transformer and cable overloads & Partly & No \\
Brockmann et al. \cite{r48} & Norway & RTP & Fleet coincidence factor & Yes & No \\
Brinkel and van Montfoort \cite{r49} & Netherlands & RTP & Grid expansion cost (aggregate peak) & Yes & No \\
Winzer and Hensler-Ludwig \cite{r50} & Switzerland & TOU & Aggregate peak, 300 households & Yes & No \\
Iftikhar et al. \cite{r51} & Pakistan & RTP & Transformer loading & Partly & No \\
Ademulegun et al. \cite{r52} & UK & TOU & Transformer upgrade need & No & No \\
Geske et al. \cite{r53} & Luxembourg & RTP & Transformer overload hours & Yes & No \\
Lilienkamp et al. \cite{r54} & Germany & RTP & Transformer loading; expansion need & Yes; smaller with price feedback & Yes \\
Qu et al. \cite{r55} & UK & Other & Aggregate EV load, 1,000 EVs & Partly & No \\
\bottomrule
\end{tabular}
\end{table*}
\begin{table}[!htbp]
\caption{Empirical studies on price-responsive electric vehicle charging and local grid stress (details in Table~\ref{tab:A2})}\label{tab:2}
\centering\scriptsize\setlength{\tabcolsep}{2pt}
\begin{tabular}{@{}>{\raggedright\arraybackslash}p{0.250\linewidth}>{\raggedright\arraybackslash}p{0.140\linewidth}>{\raggedright\arraybackslash}p{0.090\linewidth}>{\raggedright\arraybackslash}p{0.280\linewidth}>{\raggedright\arraybackslash}p{0.170\linewidth}@{}}
\toprule
\textbf{Study} & \textbf{Country} & \textbf{Signal} & \textbf{Grid metric} & \textbf{New peak?} \\
\midrule
Van Montfoort et al. \cite{r56} & Netherlands & Other & Household peak; aggregate peak, 10 households & Yes (TOU variant) \\
Brylle et al. \cite{r57} & Denmark & RTP & Transformer loading; exceedance events & Yes \\
Bailey et al. \cite{r58} & Canada & TOU & Transformer capacity violations & Yes \\
Becker et al. \cite{r59} & Belgium & TOU & Feeder peak; timing of the peak & Yes \\
Samende et al. \cite{r60} & UK & TOU & Transformer peak; voltage & Yes \\
Brinkel et al. \cite{r61} & Netherlands & RTP & Transformer congestion time & No \\
\bottomrule
\end{tabular}
\end{table}
In reviewing the publications, I focus on two questions: Do the studies find a new (rebound) peak created by the synchronized response of vehicles? And do they model the wholesale price feedback?

\begin{figure*}[!tp]
\centering
\includegraphics[width=\textwidth]{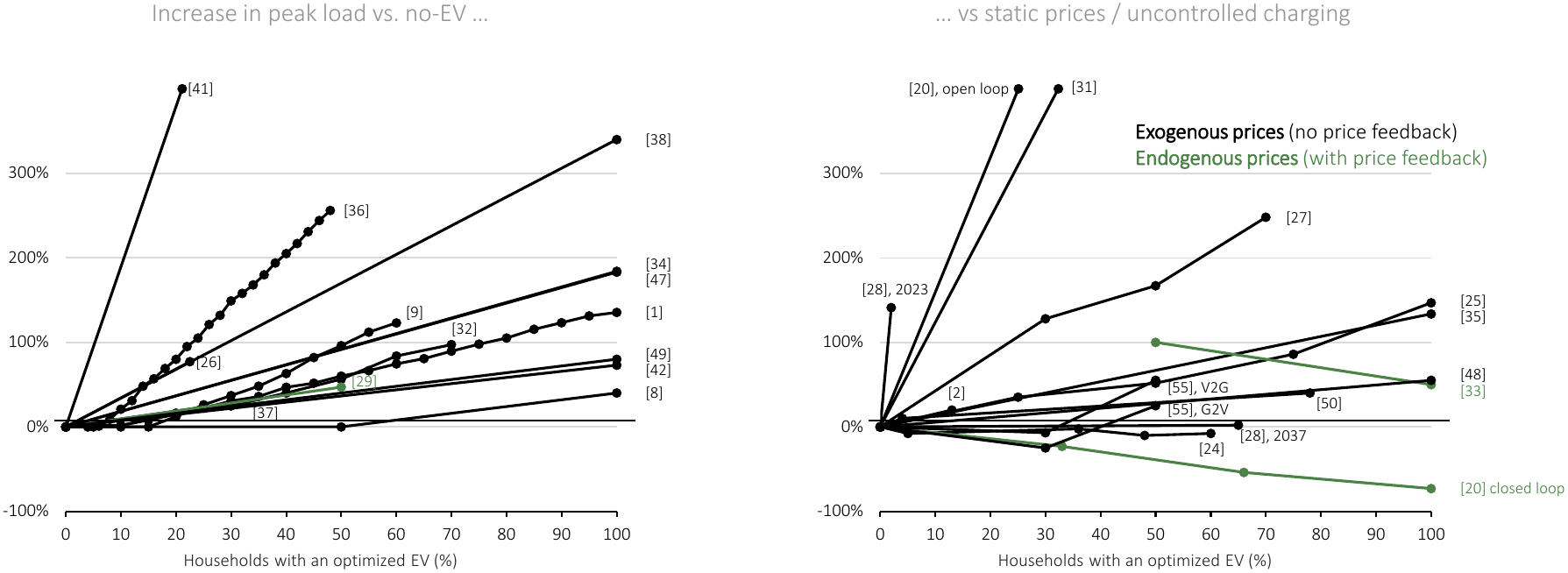}
\caption{Local grid stress (metric used in the respective study) for increasing penetration of price-responsive electric vehicles compared to no-EV (left) or uncontrolled charging (right).}\label{fig:1}
\end{figure*}

Almost every study finds a new peak (Fig.~\ref{fig:1}): charging concentrates in the hour with the lowest price, or at the start of the low-price period. According to the studies, not more than 20\,\% EV ownership is needed to create rebound peaks, often considerably less. The rebound peaks these studies report are large, even dramatic. The majority of studies find an import peak that doubles in size compared to the no-EV or uncontrolled-charging benchmarks. Some studies report three- to fivefold increases. Given that grid costs roughly scale with peak load, this would imply prohibitively expensive network expansion.

The empirical studies account for the price feedback by construction, to the extent that today's retailers anticipate the behavior of price-responsive vehicles in their procurement decisions. Anecdotal evidence suggests they do not. Also, empirical studies are limited to the low EV penetration rates observed in most countries.

Of the 38 simulation studies, only four model the wholesale price feedback: Verzijlbergh et al. \cite{r29}, Salah et al. \cite{r33}, Flath and Gottwalt \cite{r20} and Lilienkamp et al. \cite{r54}. All four papers find lower peaks than the studies with exogenous prices at comparable penetration rates, and one of them finds the peak reduced by optimized EVs.

\section{Wholesale price feedback}\label{sec:3}
If electricity consumers face time-varying prices, they shift consumption towards low-price hours. The additional demand pushes up the wholesale price in those hours, which dampens the incentive to shift. This section explains how this feedback works for vehicle charging under real-time pricing.

\subsection{Real-time pricing of vehicle charging}\label{sec:3.1}
In unbundled electricity markets, the retail price has an energy component, which the retailer derives from the wholesale market, and a regulated network component. Under real-time pricing the energy component follows the hourly or sub-hourly day-ahead price, possibly with a mark-up; time-of-use tariffs pass it on in a coarser form. This paper is about the energy component, because the feedback it studies exists in wholesale markets, not in network tariffs.

A vehicle owner with a smart charger and a real-time tariff minimizes the charging bill. Charging is moved into the cheapest hours in which the car is at home, and with a bidirectional charger the car supplies the house or the grid when prices are high (vehicle-to-home, vehicle-to-grid). The same logic applies to flexibility contracts, under which the retailer dispatches the vehicle on behalf of the owner against the wholesale price in return for free or discounted charging, such as Intelligent Octopus Go and Charge Anytime in the UK \cite{r62,r63} or the Jedlix smart charging service in the Netherlands and France \cite{r64}.

\subsection{Price feedback}\label{sec:3.2}
The wholesale price is the marginal cost of serving residual load, i.e., national load net of wind and solar generation. A fleet of vehicles that charges in the cheapest hours adds load in those hours, lifts the price and reduces the price difference that triggered the charging. Prices and charging are therefore jointly determined. In equilibrium, two conditions hold at once: wholesale prices clear the market including vehicle charging and discharging, and vehicle dispatch is optimal at those prices, so that no owner has an incentive to deviate.

In a day-ahead market the feedback works through retailers, who anticipate the response of their customers and submit price-sensitive bids, or who adjust their position on intraday and balancing markets. The model in Section~\ref{sec:4.3} does not distinguish between these channels: the price function is a reduced form of the resulting equilibrium, under the assumption that retailers anticipate the aggregate response correctly.

Flath and Gottwalt \cite{r20} distinguish open-loop prices, which are posted and do not respond to the load they trigger, from closed-loop prices, which are updated with that response. This paper uses the terms exogenous and endogenous prices in the same sense. Hirth \cite{r65} discusses the conditions under which a market equilibrium exists in more detail.

\section{Methodology}\label{sec:4}
To assess the relevance of the price feedback, a fleet of home-charged electric vehicles is optimized under different assumptions -- exogenous vs. endogenous prices. This section introduces the modeling approach and parameters.

\subsection{Measures of grid stress}\label{sec:4.1}
What matters for the costs of upgrading low-voltage grids is the peak loading of the transformer, i.e., the coincident peak of the 100 households. I measure it as the mean load in the highest 1\,\% of hours of the year, separately for imports and exports, and report it relative to the case without price-optimized vehicles. The mean of the top 1\,\% is preferred over the single highest hour because transformers tolerate short overloads and the single hour is noisy. The highest hour and the top 10\,\% of hours are reported as robustness checks (Section~\ref{sec:6.2}).

Vehicles add electricity consumption, and a higher peak is the normal consequence of higher consumption. I therefore report the coincident peak adjusted for consumption growth. Consumption is the load of all households plus the grid electricity needed for driving, i.e., home charging divided by the charging efficiency. A positive adjusted change to the no-EV peak hence indicates a disproportionate increase, one that exceeds the growth in annual consumption. If the peak rises in proportion to consumption, the adjusted change is zero. The export peak is adjusted in the same way: if all flows grow in proportion to consumption, there are no new peaks in either direction. In my model, 50\,\% EV penetration increases consumption by 18\,\%. Fig.~\ref{fig:A5} in the Appendix shows the unadjusted peaks.

\subsection{Vehicles}\label{sec:4.2}
I simulate load, solar generation and vehicle charging of 100 single-family households in hourly resolution for one year (equations in Appendix~\ref{sec:A2}). Their aggregate load corresponds to the loading of the low-voltage transformer that serves them. Load profiles are synthetic for 2025: a base load plus stochastic appliance events with the daily shape of the German standard load profile for residential consumers (BDEW H0), coupled to the weather-driven daily residual of German grid load. One in five households has an air-source heat pump, and one in five has an electric vehicle with uncontrolled charging. These 20 uncontrolled vehicles are part of the load in all scenarios. Half of the households own a PV system of 8\,kWp (ERA5 irradiance for 2025 near Berlin; east, south, west and flat orientation). Without price-optimized vehicles, the coincident import peak (top 1\,\% of hours) is 143\,kW on winter evenings, and the export peak is 183\,kW on sunny days in June. Table~\ref{tab:3} summarizes the households; Fig.~\ref{fig:A1} in the Appendix shows individual and aggregate load during the week of the annual import peak.

Price-optimized vehicles are added at random to households that do not yet own a vehicle; households own at most one electric vehicle. Their share is varied from zero to 50\,\% of all households in the base case and to 80\,\% in a sensitivity run. Each vehicle has a battery of 60\,kWh and charges at 11\,kW. Driving profiles are individual: trips, departure and return times and distances are drawn for each household, with 76\,\% commuters and a median annual mileage of 14,000\,km (5,100--30,200\,km). The car is plugged in whenever it is at home, which is 70\,\% of all hours on average. Part of the consumption is charged away from home; home charging amounts to 1.84\,MWh per car and year (0.5--4.2\,MWh).

The vehicle must have a state of charge of at least 80\,\% at each departure, and the state of charge may not fall below 20\,\% while the car is at home. Between arrival and departure, the car battery can be dispatched freely: it charges from the grid or from its own PV, supplies the house (vehicle-to-home) or injects into the grid (vehicle-to-grid) with a round-trip efficiency of 90\,\% and cycling (degradation) costs of 20\,EUR/MWh on discharge. Sensitivity runs restrict the vehicle to charging only, require a full charge at departure, and vary battery capacity, charging power and cycling cost.

All households pay a spot tariff: the hourly wholesale price plus a volumetric network charge and taxes of 10\,ct/kWh on withdrawals from the grid. Injections earn the wholesale price. For electricity that is drawn from the grid and re-injected, the network charge is refunded; losses are not. Each household minimizes its annual electricity bill under perfect foresight; for numerical reasons the problem is solved for each month separately. The decision variables are vehicle charging and discharging in each hour. Three motives drive the vehicle dispatch: the energy needed for driving, self-consumption of own solar generation, which avoids the 10\,ct/kWh of network charges and taxes, and arbitrage between low and high wholesale prices.

\begin{table}[!htbp]
\caption{Household and vehicle characteristics}\label{tab:3}
\centering\scriptsize\setlength{\tabcolsep}{2pt}
\begin{tabular}{@{}>{\raggedright\arraybackslash}p{0.340\linewidth}>{\raggedright\arraybackslash}p{0.210\linewidth}>{\raggedright\arraybackslash}p{0.120\linewidth}>{\raggedright\arraybackslash}p{0.120\linewidth}>{\raggedright\arraybackslash}p{0.120\linewidth}@{}}
\toprule
\textbf{Statistic} & \textbf{Households} & \textbf{Min} & \textbf{Mean} & \textbf{Max} \\
\midrule
Annual consumption, kWh & All 100 & 1,378 & 5,428 & 15,331 \\
Annual PV generation, kWh & 50 with PV & 6,184 & 7,111 & 8,458 \\
Highest hourly withdrawal, kW & All 100 & 3.5 & 7.2 & 17.7 \\
Highest hourly injection, kW & 50 with PV & 5.0 & 5.3 & 6.0 \\
Annual mileage, km & Optimized EVs & 5,053 & 14,138 & 30,230 \\
Home charging, kWh per year & Optimized EVs & 469 & 1,839 & 4,211 \\
\bottomrule
\end{tabular}
\end{table}
\subsection{Market modeling and price feedback}\label{sec:4.3}
The core idea of this paper is that vehicle dispatch (charging and discharging) affects wholesale electricity prices, which in turn change optimal dispatch. Equilibrium prices are the set of hourly prices that (a) are consistent with dispatch and (b) for which dispatch is optimal. The price model follows \cite{r65}: wholesale prices $p_{t}$ are a function of the residual load after flexibility (including EVs) $R_{t}$ plus an otherwise unexplained residual $\varepsilon_{t}$:

\begin{equation}
p_{t} = f(R_{t}) + \varepsilon_{t}
\tag{1}\label{eq:1}
\end{equation}
Residual load after flexibility is national load $L_{t}$ minus wind $W_{t}$ and solar $S_{t}$ generation plus the daily mean of net exports and net pumped-storage consumption ${\overline{F}}_{t}$ minus storage and vehicle dispatch $E_{t}$:

\begin{equation}
R_{t} = L_{t} - W_{t} - S_{t} + {\overline{F}}_{t} - E_{t}
\tag{2}\label{eq:2}
\end{equation}
Power system flexibility such as pumped hydro storage and imports/exports is approximated by utility-scale energy storage (superscript $u$). $E_{t}$ is the discharging $d$ minus charging $c$ of utility-scale storage and of the vehicle batteries (superscript $EV$) across $N$ identical low-voltage grids:

\begin{equation}
E_{t} = \left( d_{t}^{u} - c_{t}^{u} \right) + N \cdot \sum_{i = 1}^{H}\left( d_{i,t}^{EV} - c_{i,t}^{EV} \right)
\tag{3}\label{eq:3}
\end{equation}
The wholesale price is the marginal cost of the last thermal plant needed to serve $R_{t}$; it rises with $R_{t}$ along the merit order. My preferred specification is a cubic price function:

\begin{equation}
f(R_{t}) = a \cdot {R_{t}}^{3}
\tag{4}\label{eq:4}
\end{equation}
where $a$ is a parameter calibrated to historical prices. For robustness, I use a quadratic function:

\begin{equation}
f(R_{t}) = a_{2}\, \cdot \operatorname{sgn}(R_{t})\,|R_{t}|^{2}
\tag{5}\label{eq:5}
\end{equation}
and a piecewise linear function with ten segments at the deciles of thermal residual load.

The parameters $a$ and $a_{2}$ and the slopes of the piecewise linear function are estimated by regressing hourly German 2025 day-ahead prices on thermal residual load with day fixed effects. Thermal residual load is national residual load plus net exports and pumped-storage consumption minus pumped-storage generation (ENTSO-E). The residual between the observed and the fitted price is added as an exogenous hourly term, held fixed across all penetrations. Utility-scale batteries, pumped hydro, trade and other flexibility options are modeled as a utility-scale storage of 10\,GW and 60\,GWh with a cycling cost of 20\,EUR/MWh per direction.

Under exogenous prices, households optimize against the observed 2025 day-ahead prices, which are held fixed. Under endogenous prices, the wholesale price follows Eqs. (1) to (4) with the dispatch of the fleet, and the equilibrium is computed as one linear program per month (Appendix~\ref{sec:A2}).

The number of identical grids N determines how strongly the local fleet feeds back on the national price. In the base case, N = 190,000: 13.5 million dwellings in single-family and 5.5 million in two-family houses in Germany at the end of 2024 \cite{r66}, divided by 100 households per grid. At 50\,\% penetration this corresponds to 9.5 million price-optimized vehicles with 105\,GW of charging power and 570\,GWh of storage, in addition to 3.8 million uncontrolled vehicles.

\section{Results}\label{sec:5}
This section presents the dispatch of the vehicle fleet and the coincident peak under exogenous and endogenous prices, followed by variations of vehicle and charging parameters, charging constraints, local concentration, and renewables and network charges.

\subsection{EV charging and discharging patterns}\label{sec:5.1}
\begin{figure*}[!tp]
\centering
\includegraphics[width=\textwidth]{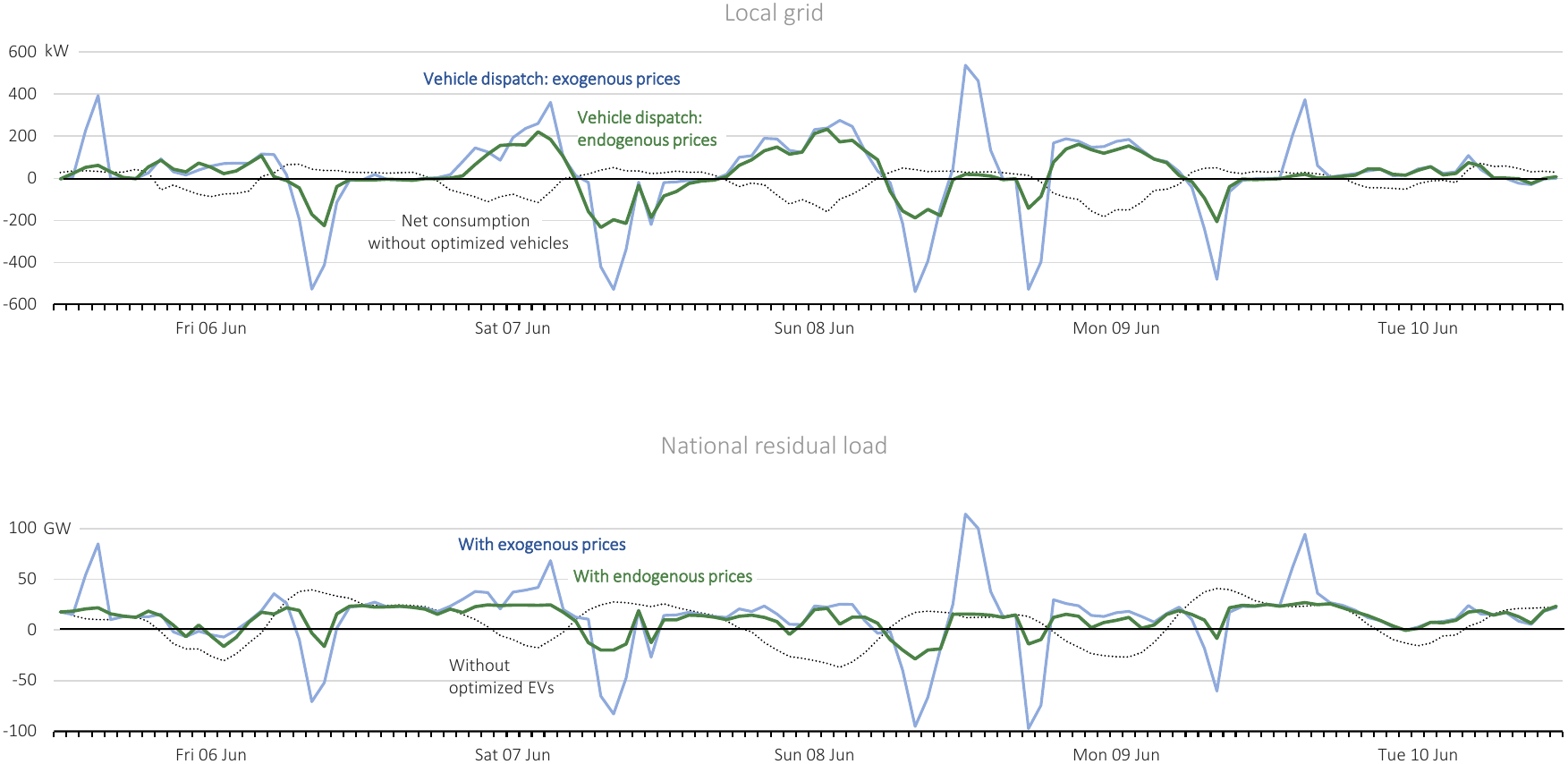}
\caption{Vehicle dispatch and resulting national residual load under exogenous and endogenous prices during five days.}\label{fig:2}
\end{figure*}

Fig.~\ref{fig:2} shows the dispatch of the 50 price-optimized vehicles (top) during five days under exogenous (blue) and endogenous (green) prices. The bottom panel shows national residual load with and without optimized EV charging. Under exogenous prices the fleet charges in bursts: in the cheapest hours, every car that is at home charges at full power, and in the most expensive hours every car discharges. Under endogenous prices, charging is spread across the low-price hours, and national residual load after flexibility is flatter than under exogenous prices.

Two mechanisms reduce fleet charging power. First, when the fleet charges into the cheapest hours, it raises the price in those hours until the price is flat across the charging window; each car then charges at lower power for more hours. Second, the feedback desynchronizes the vehicles, because the compressed price spread makes arbitrage less profitable. The share of charged energy that is discharged again falls from more than two thirds to about half. Charging patterns are more strongly shaped by individual driving profiles and PV generation, which differ across households.

\begin{figure*}[!tp]
\centering
\includegraphics[width=\textwidth]{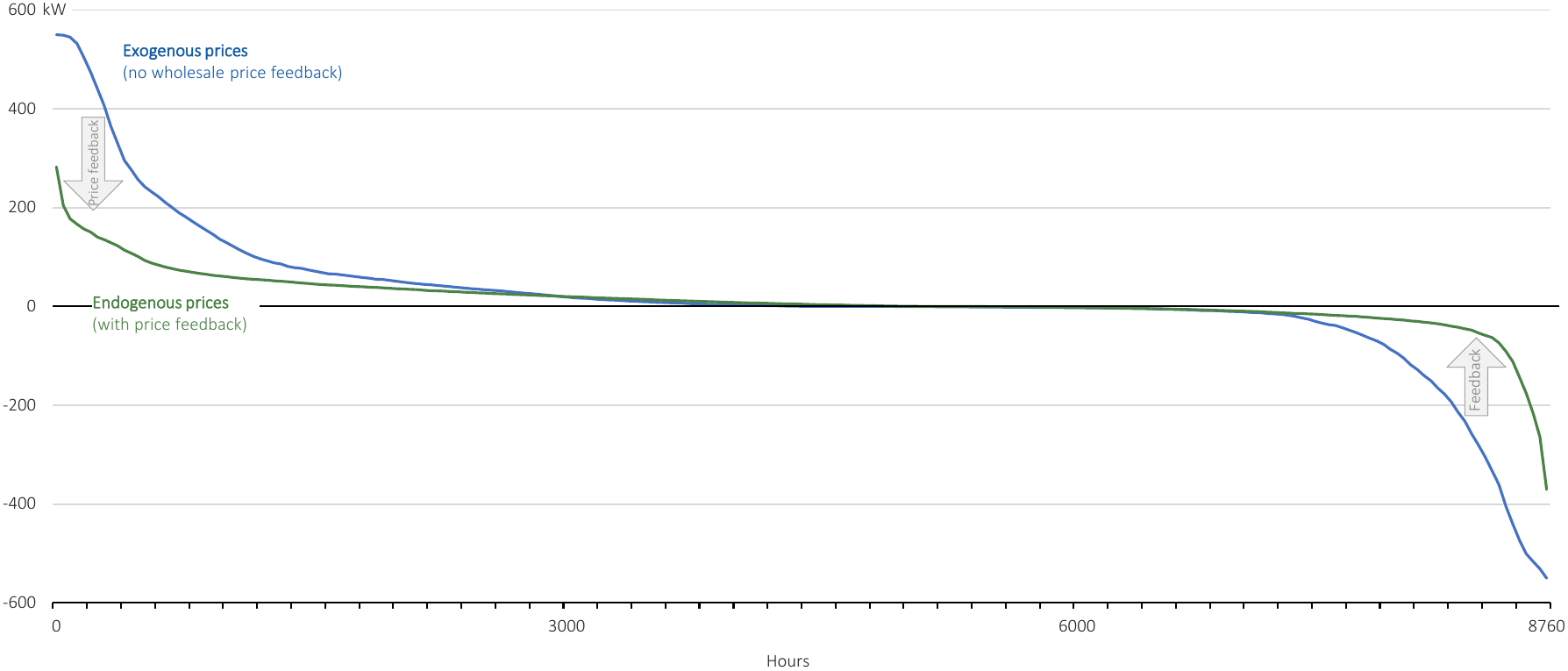}
\caption{Charging and discharging of the local vehicle fleet, ordered by size (duration curves).}\label{fig:3}
\end{figure*}

Fig.~\ref{fig:3} shows the resulting duration curves of fleet charging and discharging. Under exogenous prices, the fleet reaches its full rated charging power in dozens of hours of the year and its full rated discharging power in several hours. Under endogenous prices this never happens; fleet charging peaks at about half the rated power.

\subsection{Coincident peak}\label{sec:5.2}
\begin{figure*}[!tp]
\centering
\includegraphics[width=\textwidth]{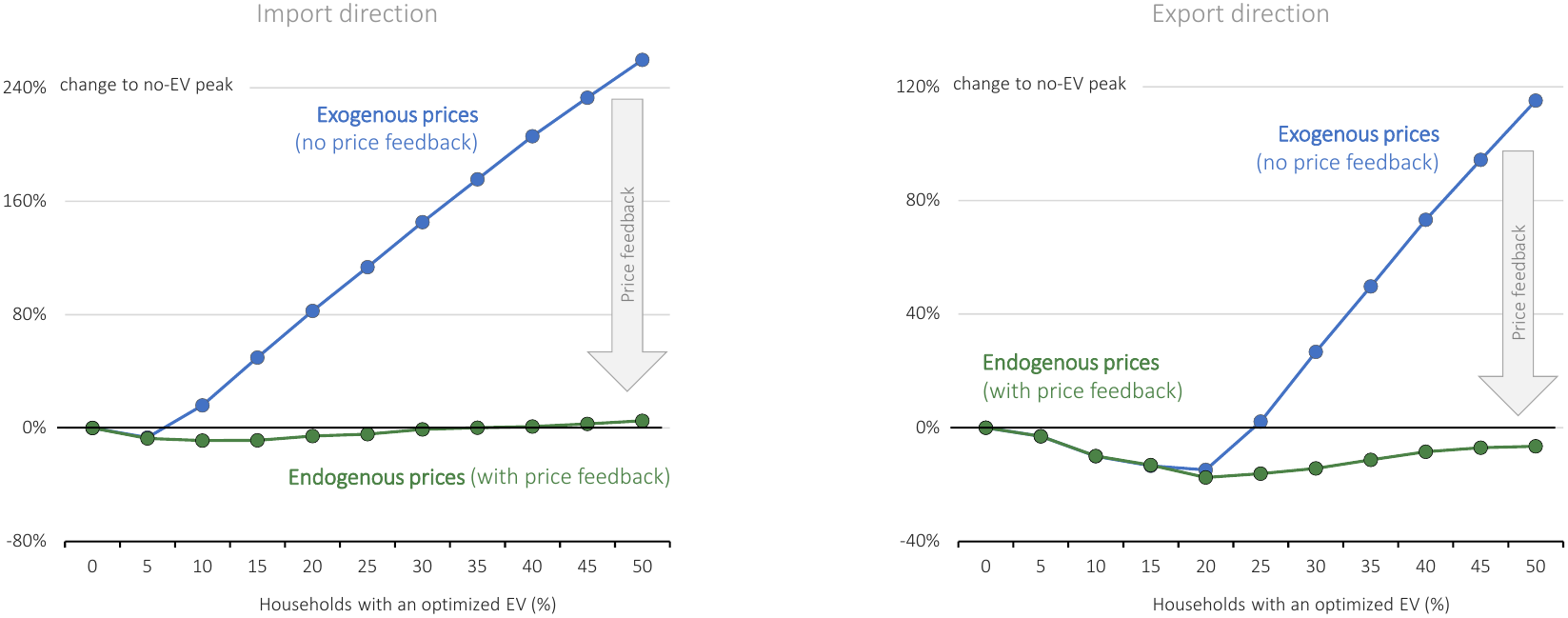}
\caption{Coincident peak for exogenous and endogenous prices, import and export direction.}\label{fig:4}
\end{figure*}

Fig.~\ref{fig:4} shows the coincident peak for a growing share of households with a price-optimized vehicle, in the import (left) and export (right) direction. With few vehicles, the peaks fall slightly under both price assumptions: vehicles charge when local load and prices are low, and they discharge in the evening, when both are high. This is no longer true once the fleet of optimized cars expands. Under exogenous prices, the import peak exceeds its no-EV level once 5--10\,\% of households own an optimized vehicle, and it rises linearly thereafter. At 50\,\% penetration, it is 260\,\% above the no-EV peak, a rather dramatic result. This increase already accounts for the 18\,\% increase in annual electricity consumption caused by EVs. If grid costs scale with peak load, this implies 260\,\% higher low-voltage grid costs per kWh consumed. The coincidence factor, calculated as the coincident load in the top 1\,\% of hours divided by the sum of individual households' top 1\,\% of hours, increases from 0.30 to 0.66.

This is the rebound peak that the literature finds: all cars charge perfectly synchronized at full power in the same hours, causing new peaks in what used to be off-peak periods (Fig.~\ref{fig:A2} in the Appendix). Where old import peaks occurred on winter evenings, new import peaks emerge in winter nights, when synchronized charging meets non-negligible household demand. The export peak exceeds its no-EV level from about 25\,\% penetration and reaches +115\,\% at full penetration, because all cars discharge into the grid in the same high-price hours.

Under endogenous prices, the import peak stays below its no-EV level up to about 35\,\% penetration and rises to +5\,\% at full penetration; the export peak stays below its no-EV level at all penetrations and ends at -6\,\%. The price feedback thus removes almost the entire increase in both directions. Without the adjustment for consumption growth, the endogenous import peak rises by about a quarter at full penetration (Fig.~\ref{fig:A5} in the Appendix).

While not identical, these results are consistent with the four previous studies that model the price feedback.

Verzijlbergh et al. \cite{r29} simulate an aggregator that minimizes the charging cost of about 500 vehicles on one Dutch distribution cable serving 1,000 households, without vehicle-to-grid. The wholesale price is an exogenous price path plus a term linear in the vehicle load. They find that cost-minimizing charging raises the cable peak by 47\,\% with 15\,GW of wind, but creates no new peak with 2\,GW. Their peak is not adjusted for consumption growth: uncontrolled charging alone raises it by 35\,\%.

Salah et al. \cite{r33} study 49 high- to medium-voltage substations of a Swiss utility with load forecasts for 2040 and home charging at 11\,kW, without vehicle-to-grid. The spot price is a linear function of Swiss residual load, including the charging of all vehicles in the country. They find fewer substation overloads under a dynamic tariff than under a flat tariff with low renewable shares, but more with medium and high shares, because larger price spreads concentrate charging despite the feedback. Their overloads from 16\,\% penetration onward reflect a threshold metric: they count substations above capacity on a peak-day forecast for 2040 that is already close to it.

Lilienkamp et al. \cite{r54} embed 16,191 representative German low-voltage grids, each with a transformer limit, into a European dispatch model in which wholesale prices are the duals of the market-clearing program. With exogenous prices, the peak of the national charging profile of 15 million vehicles is 75\,\% higher than with endogenous prices. The feedback attenuates rather than removes the peak: even with endogenous prices, charging adds up to 40\,GW in single hours, and up to 7.8\,GW must be curtailed to respect transformer limits.

Flath and Gottwalt \cite{r20} simulate a microgrid of 900 households with vehicles and wind power; their metric is the peak reserve requirement. Under an open-loop price, linear in local net wind, this peak rises more than tenfold relative to a flat tariff. They find that a closed-loop price lowers the peak well below that under a flat tariff. Their price follows the net load of the same microgrid, so the feedback acts directly on the load that forms the peak -- this resembles distribution-level nodal pricing. Here, the national price reaches each local grid only through the correlation of local and national load.

\subsection{Vehicle and charging parameters}\label{sec:5.3}
\begin{figure*}[!tp]
\centering
\includegraphics[width=\textwidth]{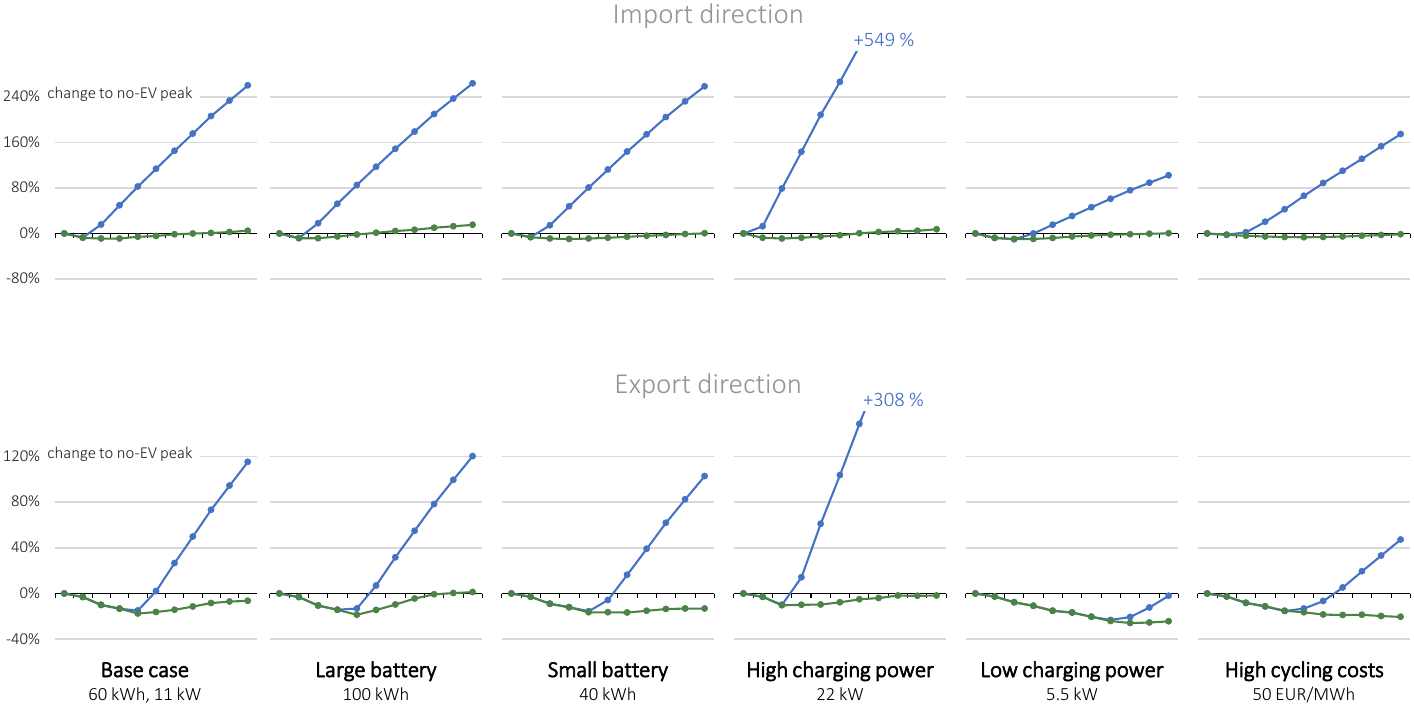}
\caption{Coincident peak for exogenous and endogenous prices: vehicle and charging parameters.}\label{fig:5}
\end{figure*}

This section varies EV parameters: a larger and a smaller battery, double and half the charging power, higher cycling costs, and price-optimized vehicles in all households without an uncontrolled vehicle. Fig.~\ref{fig:5} shows, for each vehicle parameter separately, the coincident peak under exogenous (blue) and endogenous (green) prices; Table~\ref{tab:A4} in the Appendix reports the values at full penetration for all model runs.

In every variation the price feedback cuts the import peak at full penetration by at least half, in most by much more. Charging power is the parameter that matters most. At double the charging power, the exogenous import peak rises more than sixfold, while the endogenous import peak changes little. Battery capacity matters less: the endogenous import peak rises somewhat more with the larger battery and not at all with the smaller one. Higher cycling costs reduce discharging and with it both peaks. With vehicles in all households, the exogenous import peak rises more than fivefold, the endogenous import peak only moderately.

The cases that show the most extreme rebound peaks, high power and high penetration, are of course also the cases where optimized charging has the largest benefits for the wider power system, in particular reducing generation costs, supplying balancing reserves, and providing firm capacity for generation adequacy.

What matters most in the context of this paper is the robustness with which the wholesale market feedback eliminates rebound peaks. The worst of these cases, an increase of battery capacity to 100\,kWh per EV, results in a new peak that is just 15\,\% larger than the no-EV case, well within the margin of most low-voltage transformers.

\subsection{Charging constraints}\label{sec:5.4}
\begin{figure*}[!tp]
\centering
\includegraphics[width=\textwidth]{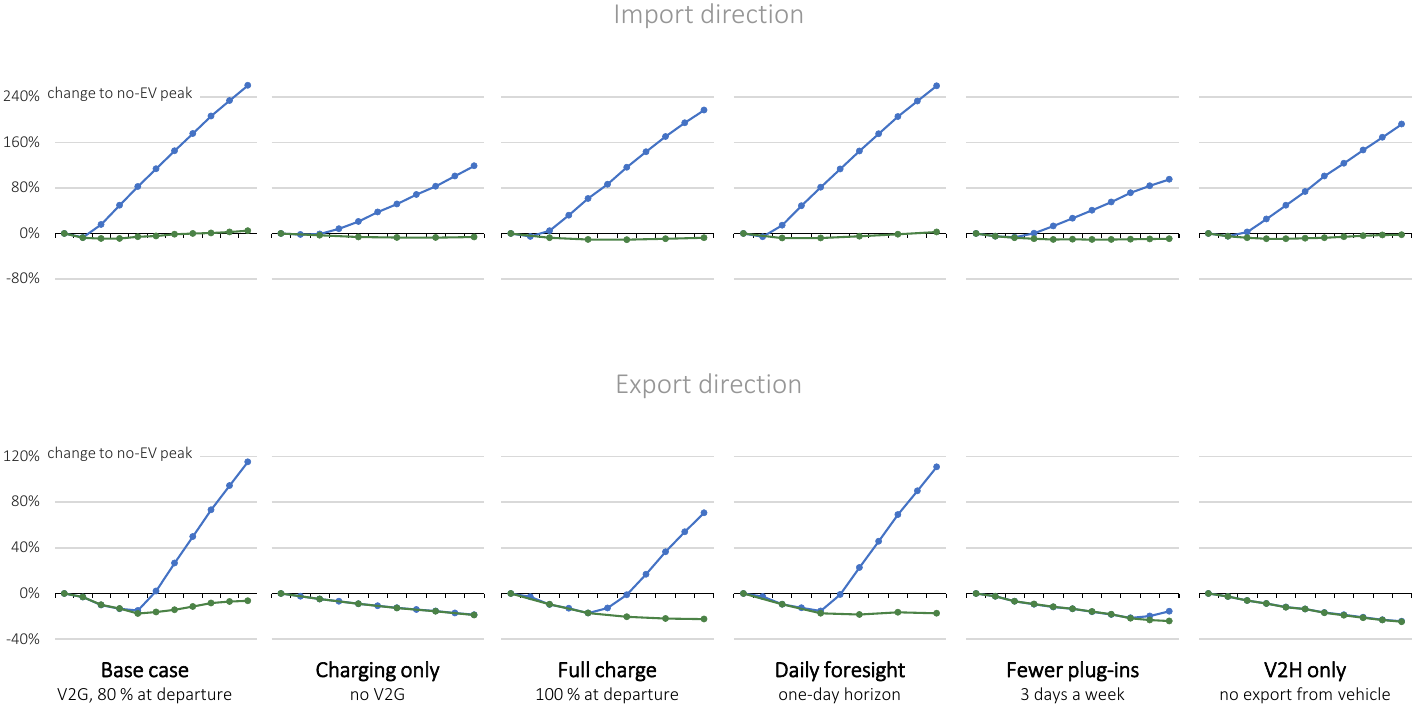}
\caption{Coincident peak for exogenous and endogenous prices: charging constraints.}\label{fig:6}
\end{figure*}

The base case allows bidirectional charging with perfect foresight within each month and requires 80\,\% state of charge at departure. Fig.~\ref{fig:6} shows results for changes to these assumptions. With ``only charging'' (no V2G), vehicles do not add to the export peak; the import peak more than doubles under exogenous prices and falls slightly under endogenous prices. Requiring a full charge at each departure reduces the arbitrage volume and lowers both peaks. A one-day optimization horizon instead of one month changes little. If each vehicle is plugged in on only three days a week, the exogenous import peak roughly doubles instead of more than tripling; under endogenous prices, it stays below its no-EV level. If the vehicle may only supply the household's own load (V2H only), the export peak falls below its no-EV level under both price regimes, while the exogenous import peak still almost triples. In all variations the endogenous import peak at full penetration stays close to or below its no-EV level.

Restricting the vehicles -- lower charging power, higher cycling costs, no vehicle-to-grid or, in the export direction, a smaller battery -- lowers the peaks under exogenous prices. It also lowers the value of the fleet to the power system: less energy arbitrage and hence lower savings in generation costs, less firm capacity for generation adequacy, and less capacity for balancing and other ancillary services.

\subsection{Local concentration}\label{sec:5.5}
The price feedback works at the level of the wholesale market, while grid stress is local. In Germany, with its uniform wholesale market, that is the entire country; elsewhere it is the bidding zone or, under locational marginal pricing, the transmission node. As long as local grids are similar in load patterns and vehicle penetration, the national price smooths local charging nevertheless. If vehicles are concentrated in few grids, e.g., affluent suburban neighborhoods, the national price responds less than the local fleet would require.

\begin{figure*}[!tp]
\centering
\includegraphics[width=\textwidth]{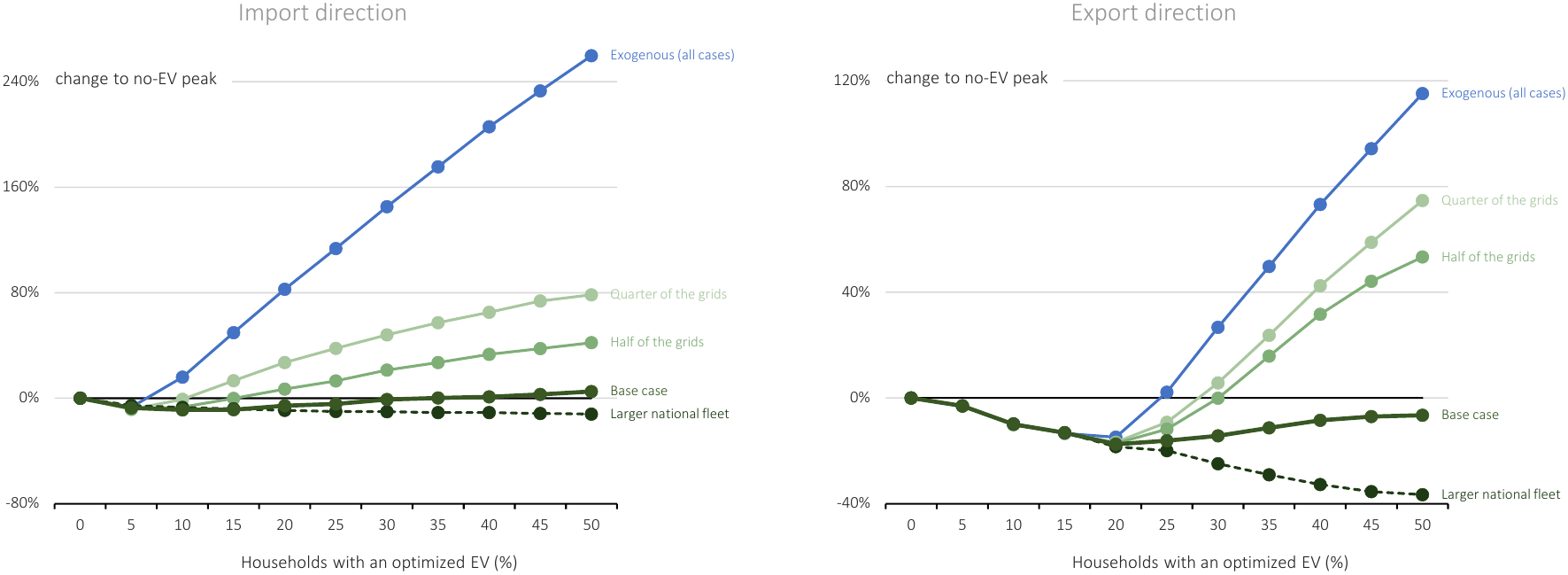}
\caption{Coincident peak for exogenous and endogenous prices: local concentration.}\label{fig:7}
\end{figure*}

In the simulation, this is captured by the number of identical grids N in Eq. (3): a quarter and half of the grids (the national fleet is a quarter and half of the base case at the same local penetration), and a larger national fleet (twice the base case). With a quarter of the grids, the endogenous import and export peaks rise by about three quarters at full penetration; with half of the grids by roughly half (Fig.~\ref{fig:7}). A larger national fleet strengthens the feedback. Both peaks then fall below their no-EV level. In the limit of a single grid, the exogenous-price case is reproduced. Utility-scale flexibility matters less: without it, the endogenous import peak rises somewhat more (Fig.~\ref{fig:A3} in the Appendix).

\subsection{Renewables}\label{sec:5.6}
\begin{figure*}[!tp]
\centering
\includegraphics[width=\textwidth]{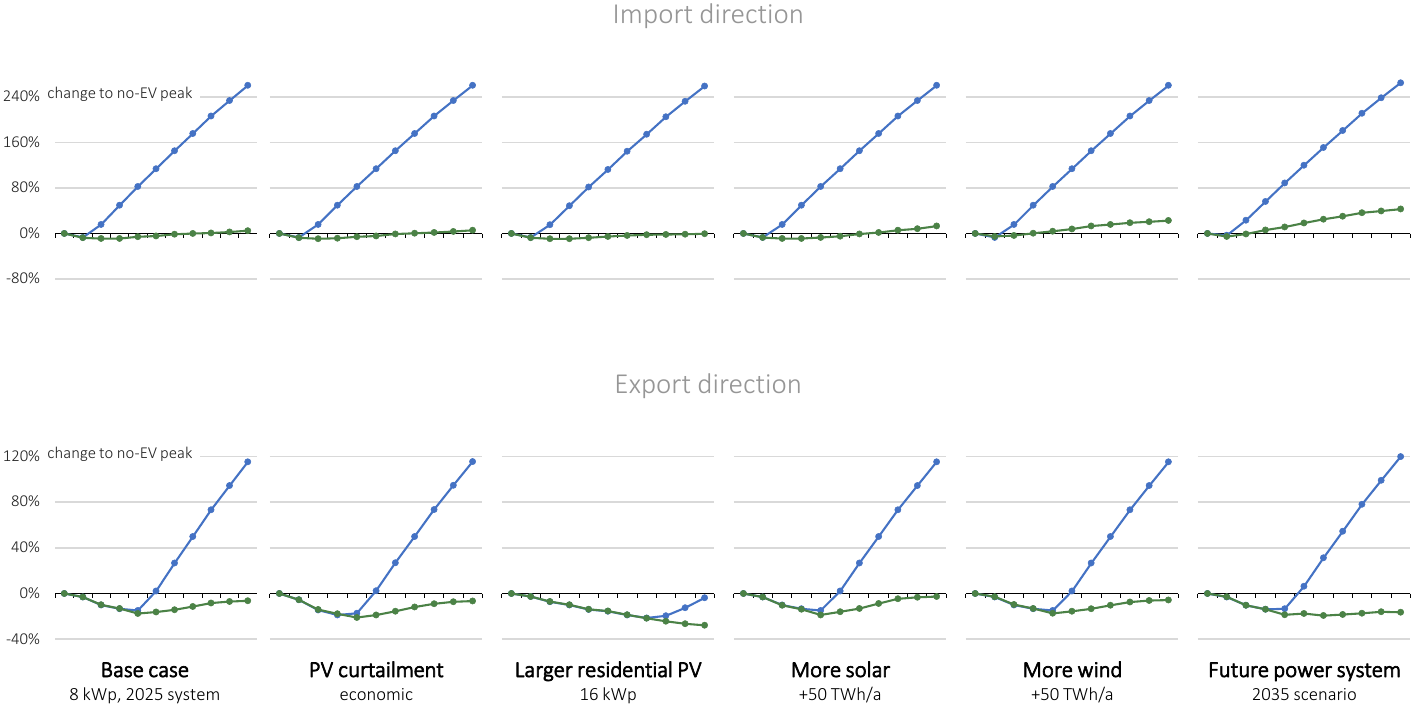}
\caption{Coincident peak for exogenous and endogenous prices: renewables.}\label{fig:8}
\end{figure*}

Wind and solar generation shape residual load and hence wholesale prices; residential solar shapes the local export peak and the self-consumption motive. Fig.~\ref{fig:8} shows variations of both. Economic curtailment of residential PV and larger residential PV systems change little. More utility-scale solar, more wind and, most of all, a 2035 scenario with much higher renewable capacity, load and utility-scale flexibility raise the endogenous import peak. More renewable supply makes low-price hours more frequent and the price function flatter around them, so that the feedback is weaker; in each case it still removes more than four fifths of the increase found under exogenous prices.

\subsection{Network charges}\label{sec:5.7}
\begin{figure*}[!tp]
\centering
\includegraphics[width=\textwidth]{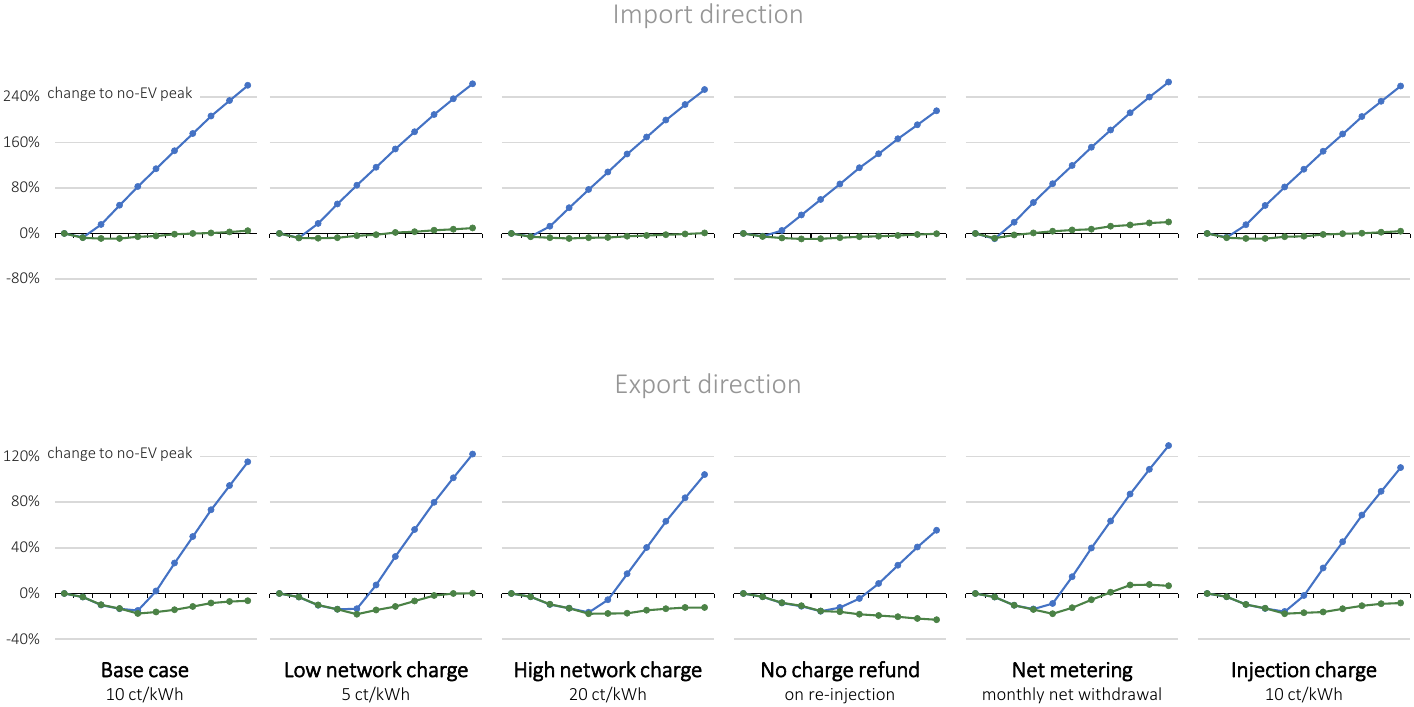}
\caption{Coincident peak for exogenous and endogenous prices: network charges.}\label{fig:9}
\end{figure*}

Halving or doubling the network charge changes the peaks only slightly (Fig.~\ref{fig:9}). Without the refund of the network charge on re-injected electricity, vehicle-to-grid arbitrage is less attractive, and both peaks fall. Two further variants were tested: net metering, under which the charge of 10\,ct/kWh applies only to each household's monthly net withdrawal, and a charge of 10\,ct/kWh also on injection. In all cases, endogenous prices remove the rebound peak to a large degree.

\begin{figure*}[!tp]
\centering
\includegraphics[width=\textwidth]{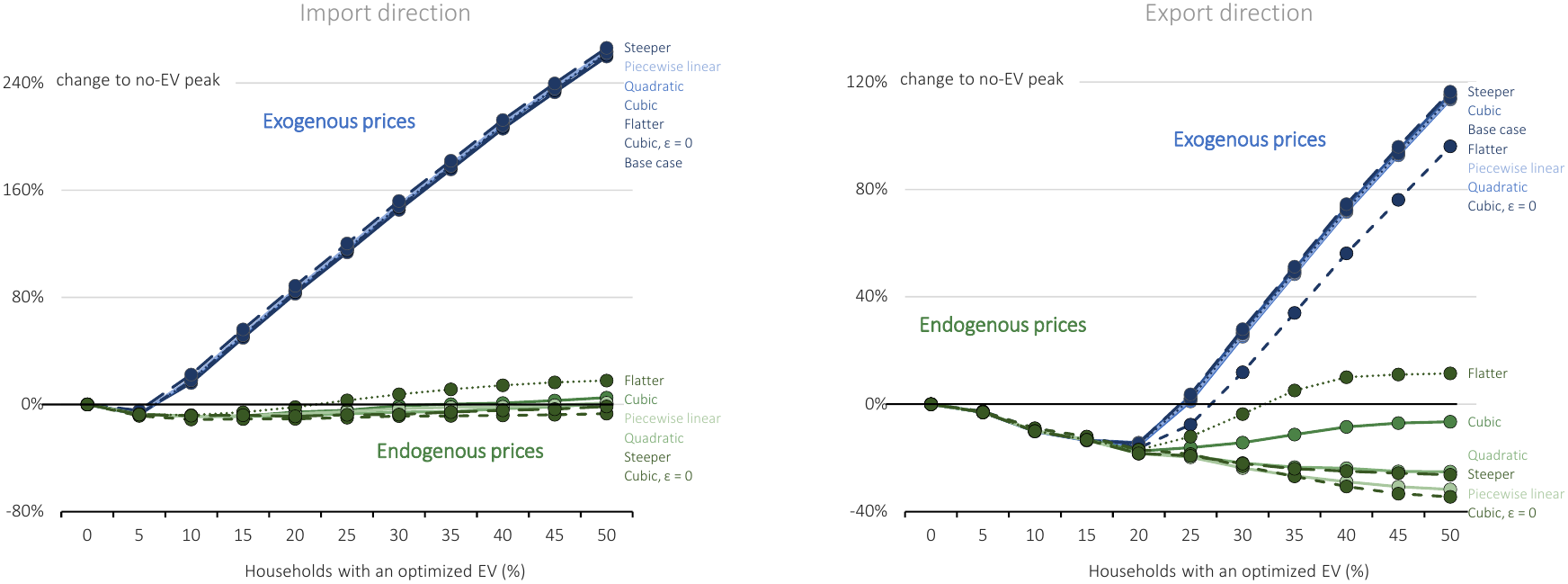}
\caption{Coincident peak for alternative price functions and different exogenous price curves.}\label{fig:10}
\end{figure*}
\begin{figure*}[!tp]
\centering
\includegraphics[width=\textwidth]{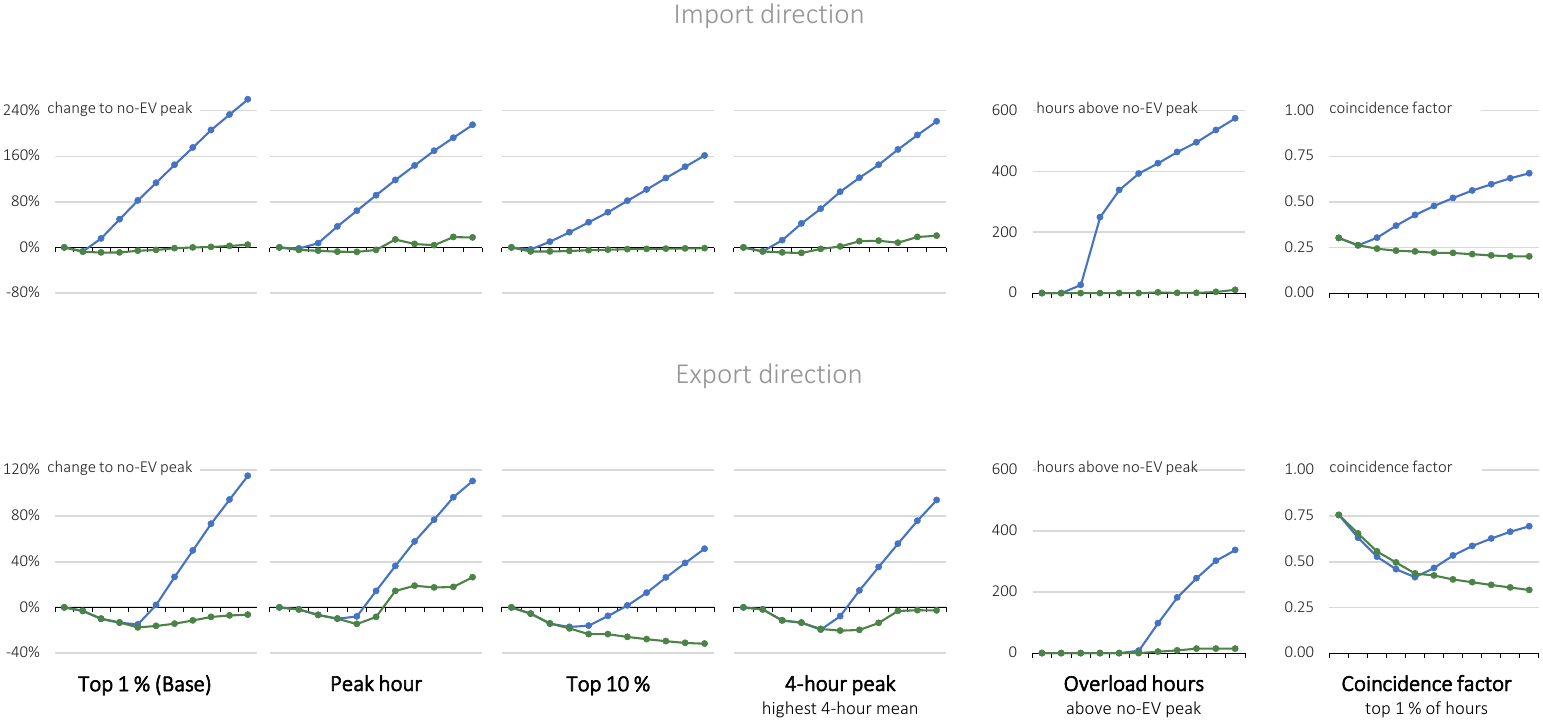}
\caption{Alternative measures of the coincident peak (maximum, top 1\,\%, top 10\,\%, 4-hour peak), overload hours and coincidence factor for exogenous and endogenous prices.}\label{fig:11}
\end{figure*}
\section{Robustness}\label{sec:6}
This section tests the robustness of the results against the functional form of the price function, the measure of the coincident peak and the household simulation procedure, and summarizes all model runs.

\subsection{Alternative price functions}\label{sec:6.1}
The core idea of the price feedback, or wholesale market equilibrium, is that spot prices are a function of residual load after flexibility (including EV optimization). The base case uses a cubic function and historical 2025 prices as the exogenous case. Here I test five alternative functions: quadratic, piecewise linear, cubic function without the exogenous price residual ($\varepsilon$ = 0), flatter cubic (calibration parameter halved) and steeper cubic (parameter doubled).

I also vary the exogenous price series: besides observed 2025 prices, I calculate the prices for each function for the no-EV case and hold them fixed. This allows a clean comparison between closed-loop and open-loop modeling.

Across all variants, the endogenous import peak at full penetration stays close to its no-EV level, whereas the exogenous import peak more than triples (Fig.~\ref{fig:10}). The endogenous export peak lies below its no-EV level in all variants except the flatter price function. The flatter price function, which weakens the feedback, gives the highest peaks; the function without the price residual, which removes the exogenous part of price volatility, the lowest.

\subsection{Alternative measures of the coincident peak}\label{sec:6.2}
My preferred measure of local grid stress is the mean load in the top 1\,\% of hours. Fig.~\ref{fig:11} presents a range of alternative measures: the load in the highest hour of the year, the mean of the top 10\,\% of hours, the peak of the four-hour moving average, the number of overload hours (hours with higher load than the no-EV case), and the coincidence factor, the coincident peak divided by the sum of the individual households' peaks (both top 1\,\%).

In the highest hour, both endogenous peaks at full penetration lie moderately above their no-EV level; in the top 10\,\% of hours, both lie below it. For every measure, the price feedback removes most of the increase found under exogenous prices.

\subsection{Household simulation assumptions}\label{sec:6.3}
The last block of Table~\ref{tab:A4} varies the simulation procedure: the draw of households (400 bootstrap replications and three independently generated networks), the number of households per grid (25 to 400, national fleet unchanged), the coupling of household load to national load, measured instead of synthetic load profiles (EKZ Zurich 2024, CKW Lucerne 2025), quarter-hourly resolution, the historical years 2023 and 2024, PV orientation, and the share of households with PV.

In every case the price feedback removes most of the increase in both directions. The endogenous import peak rises only moderately in all variations with synthetic load profiles and more with the measured profiles from Zurich and Lucerne, which have a more pronounced evening peak. With fewer PV households, the export peak rises strongly in relative terms, because the no-EV export peak is much smaller.

\subsection{Overall robustness}\label{sec:6.4}
\begin{figure}[!t]
\centering
\includegraphics[width=\columnwidth]{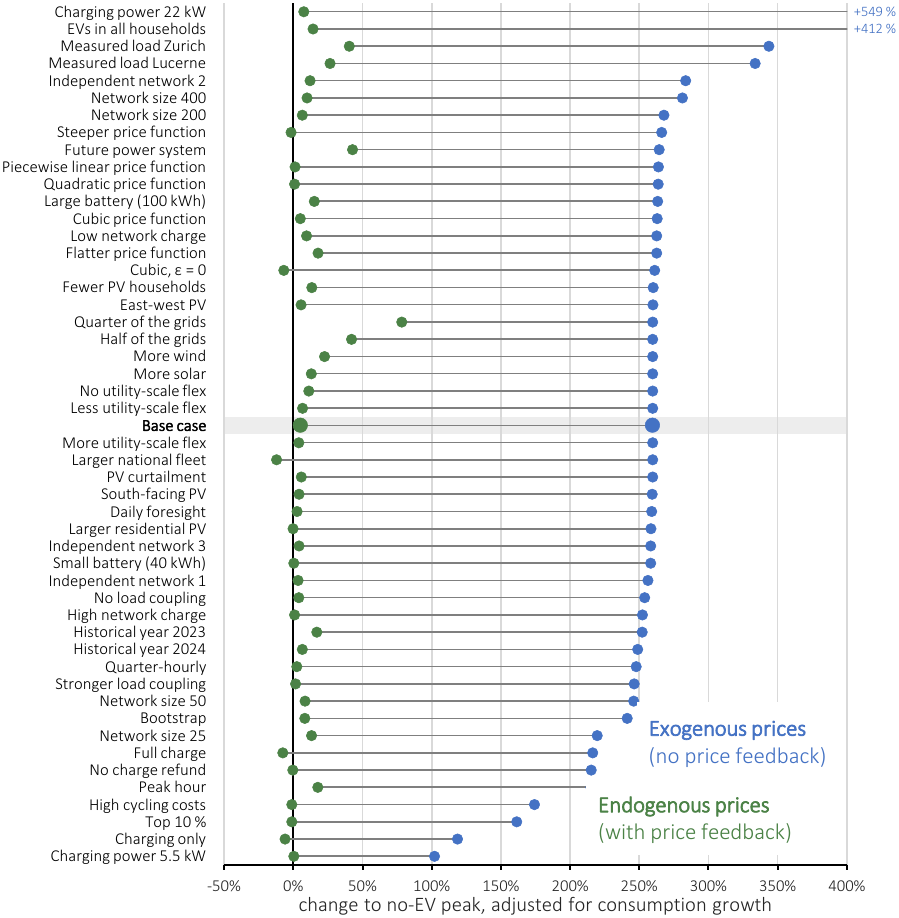}
\caption{Coincident peak at full penetration across all model runs, import direction: endogenous (green) vs. exogenous (blue) prices.}\label{fig:12}
\end{figure}

\begin{figure}[!htbp]
\centering
\includegraphics[width=\columnwidth]{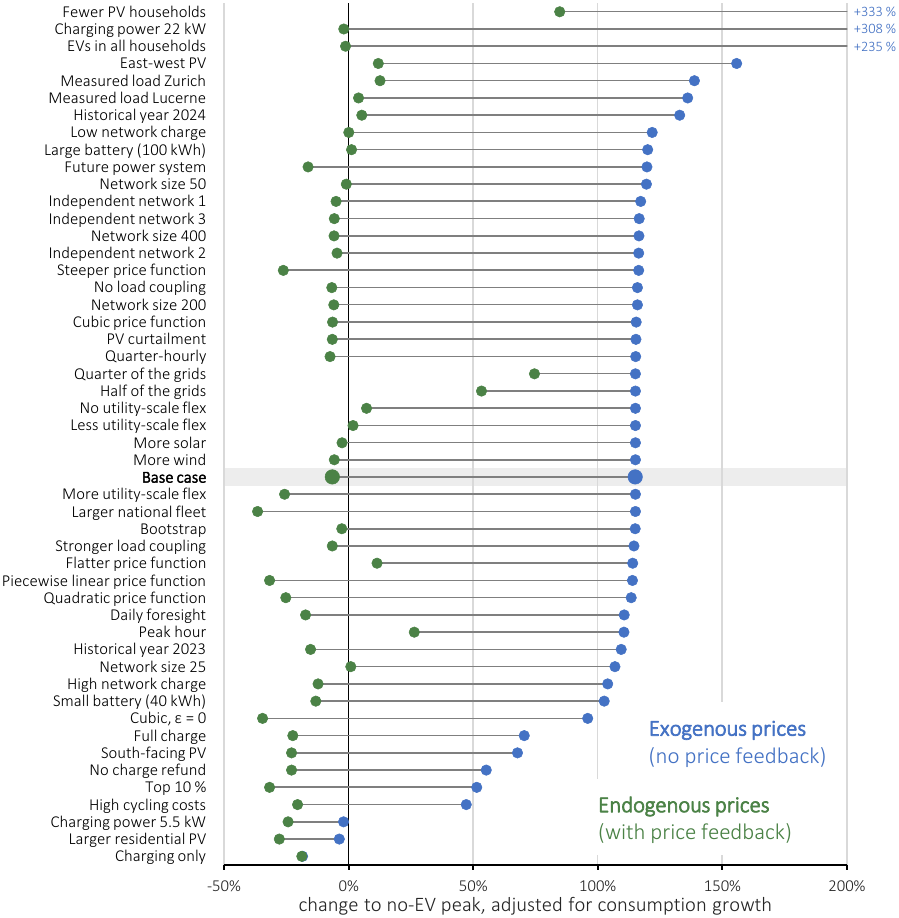}
\caption{Coincident peak at full penetration across all model runs, export direction: endogenous (green) vs. exogenous (blue) prices.}\label{fig:A4}
\end{figure}

Fig.~\ref{fig:12} summarizes the coincident import peak at full penetration under exogenous and endogenous prices for all model runs (Fig.~\ref{fig:A4} shows the export direction). Two findings stand out. First, the price feedback matters a lot. The difference between exogenous and endogenous prices amounts to at least a hundred percentage points, in every single run. In most runs more than two hundred, in a few more than three hundred.

Second, while under exogenous prices the import peak rises by multiples, under endogenous prices the increase stays modest across all runs. Except for grids in local clusters of high EV penetration, the increase in coincident peak remains below 50\,\%. In many low-voltage grids, this is likely within the remaining capacity margins of network elements. And these are the extreme scenarios -- in most model runs the increase is even less, sometimes negative.

These results suggest that price-responsive vehicles may raise local peaks roughly in proportion to the consumption they add, far less than exogenous-price simulations suggest.

\cleanbarrier
\section{Conclusions}\label{sec:7}
This paper asks whether the dramatic rebound peaks that the literature attributes to price-responsive electric vehicles persist once wholesale prices respond to the charging they trigger. Simulations suggest that under a broad range of scenarios, they do not.

With exogenous prices, the simulations presented in this paper reproduce the literature: once more than 5--10\,\% of households charge a vehicle against the real-time price, their synchronized charging creates a new peak in winter nights, and at 50\,\% penetration the coincident import peak of the local grid, adjusted for consumption growth, is 260\,\% above its no-EV level. With endogenous prices, the same fleet raises the peak by just 5\,\%. The result holds across vehicle and charging parameters, charging constraints, renewable scenarios, network charges, price functions, peak measures and simulation assumptions.

The price feedback does not rule out larger peaks: they occur where vehicle adoption is concentrated in few grids, where charging power is high, and in the single highest hour of the year. Grid operators have reason to watch local clusters and charging power; they have less reason to expect the peaks that exogenous-price simulations predict.

\newpage
For grid planning, this means that reinforcement needs derived from exogenous-price simulations overstate the capacity that price-responsive vehicles require. For regulation, it means that real-time pricing of vehicle charging is less of a threat to the distribution grid than the literature suggests. This weakens the case for dedicated instruments such as staggered time-of-use windows or limits on charging power, without removing it for grids where adoption runs ahead of the national average. The price-feedback model used here is simple enough to be added to existing simulation studies.

\cleanbarrier
\appendices
\onecolumn
\section{Studies on price-responsive electric vehicle charging}\label{sec:A1}
{\footnotesize\setlength{\tabcolsep}{3pt}
\begin{longtable}{@{}>{\raggedright\arraybackslash}p{0.121\linewidth}>{\raggedright\arraybackslash}p{0.171\linewidth}>{\raggedright\arraybackslash}p{0.128\linewidth}>{\raggedright\arraybackslash}p{0.115\linewidth}>{\raggedright\arraybackslash}p{0.299\linewidth}>{\raggedright\arraybackslash}p{0.085\linewidth}@{}}
\caption{Simulation studies on price-responsive electric vehicle charging and local grid stress}\label{tab:A1}\\
\toprule
\textbf{Study} & \textbf{Setting} & \textbf{Signal} & \textbf{Grid metric} & \textbf{Synchronized new peak?} & \textbf{Price feedback} \\
\midrule
\endfirsthead
\toprule
\textbf{Study} & \textbf{Setting} & \textbf{Signal} & \textbf{Grid metric} & \textbf{Synchronized new peak?} & \textbf{Price feedback} \\
\midrule
\endhead
Stute and Kühnbach \cite{r24} & Germany; 146-node suburban low-voltage grid; HEMS with battery, EV, heat pump & RTP: EPEX day-ahead prices of 2019, or TOU, three-tier; freely chosen by households & Household peak (withdrawal); transformer loading; overload hours & Partly. Household peaks up 11 and 15\,\% and overload hours up from 2 to 111 per year; the grid maximum falls because the tariff mix limits synchronization & No \\
Semmelmann et al. \cite{r25} & Germany; 500 single-family households, each with EV, heat pump, PV and battery, empirical load and charging data, hourly, 2019; ten representative medium- and low-voltage grids; four peak day types & RTP: day-ahead prices of 2019, against a constant price; adoption 0 to 100\,\%; four network charge designs and two feed-in schemes & Reinforcement cost (medium and low voltage) & Yes. Under volumetric network charges, reinforcement cost across the ten grids rises from 22 to 55 million EUR between zero and full adoption; some grids need reinforcement from 25\,\% adoption, at full adoption all grids but one. Segmented network charges hold the rise below 1 million EUR & No \\
Cong et al. \cite{r26} & Denmark; one real 400\,kVA feeder with 160 households; EVs, PV and home batteries; second-by-second agent-based model, 2024--2025 & RTP: Nord Pool day-ahead prices, plus a time-of-use network charge & Transformer loading; overload hours (import) & Yes. Price-driven EV charging overloads the transformer within four months of the adoption path & No \\
Lilienkamp and Namockel \cite{r27} & Germany; SimBench semi-urban grid with 12 low-voltage grids and 940 households; EVs, PV, heat pumps and storage; 30 to 70\,\% EV penetration, 2030 & RTP: quarter-hourly day-ahead prices for 2030 from a market model; TOU with three windows derived from them & Transformer overload events (import) & Yes. At 50\,\% EV penetration no overloads under the flat tariff and about 4,800 per year under TOU or RTP, attributed to herding & No \\
Lersch et al. \cite{r28} & Germany; IEEE European low-voltage test feeder with 55 buildings; PV, home batteries, EVs and heat pumps; adoption of 2023 and 2037; one summer week & RTP: German day-ahead prices of 2023; variant with a dynamic network charge & Transformer and line loading (import); voltage; reinforcement cost & Yes. In the 2023 case dynamic prices raise peak transformer loading from 29 to 70\,\% of rating, with the dynamic network charge to 91\,\% & No \\
Turk et al. \cite{r1} & Massachusetts; 400 households on one feeder; EVs with perfect foresight & TOU energy rate, two-part, plus seven network tariff designs & Feeder peak (import) & Yes. Rebound peak at the start of the off-peak period exceeds the historical evening peak from 15\,\% EV adoption; at 50\,\% adoption it is 60\,\% above the historical peak & No \\
Verzijlbergh et al. \cite{r29} & Netherlands; one distribution cable with about 1,000 households and 500 EVs & RTP: wholesale prices from a Dutch system model & Feeder peak (import) & Yes. Cost-minimizing charging raises the cable peak by 47\,\%, more than uncontrolled charging (35\,\%); with 2\,GW wind the new peak does not occur. Static time-of-use network tariffs raise it by up to 69\,\% & Yes \\
Stawska et al. \cite{r30} & Netherlands; 250 households on one 500\,kW feeder; EV aggregator & Other: historical Dutch imbalance prices; energy tariff, capacity tariff or local flexibility market & Feeder overload hours & Yes. Charging spikes whenever imbalance prices are very low, growing with penetration and charging power; an energy-based network tariff does not prevent it & No \\
Flath et al. \cite{r31} & Germany; EV fleet; residential distribution grids & Other: time-based prices, uniform price vs. area pricing with a local capacity component & Distribution grid peak (import) & Yes. Time-based prices alone produce high load spikes from simultaneous charging, independent of charging strategy and power level & No \\
Huang and Sandström \cite{r32} & Sweden; residential district with 1,321 customers on twelve 10/0.4\,kV substations; EVs with V1G and V2G at 5 to 70\,\% penetration & RTP: spot price, plus a time-of-use network tariff; also prices differentiated by substation & Transformer loading; cable loading; voltage; hosting capacity & Yes, above 20\,\% penetration. V2G lowers peaks at low penetration and raises them above it. Prices differentiated by substation lower the district peak but not the peak of the individual substation & No \\
Salah et al. \cite{r33} & Switzerland; 49 distribution substations of one regional grid with forecast 2040 peak-day load; survey-based driving profiles; home charging at 11\,kW; penetration of 16, 50 and 100\,\% & RTP: hypothetical dynamic tariff following the Swiss spot price, against a flat tariff & Substation loading; share of substations overloaded (import) & Yes, with medium and high renewable shares: overloads from 16\,\% penetration despite the price feedback. With low renewable shares fewer overloads than under the flat tariff, which overloads 6 of 49 substations at full penetration & Yes; spot price as a function of residual load, updated after each vehicle's charging decision \\
von Bonin et al. \cite{r34} & Germany; 100 metropolitan households, each with EV, PV and 11\,kW wallbox; home energy management; 2030 prices & RTP: hourly day-ahead prices for 2030 from a market simulation & Aggregate peak, 100 households (import) & Yes. Price-optimized charging raises the aggregate peak by 184\,\% over household load, direct charging by 84\,\%; the relative increase grows with the number of households & No \\
Panda et al. \cite{r35} & Netherlands; measured sessions at about 1,200 public charging points, 2022; aggregation of 1 to 1,024 points & RTP: Dutch day-ahead prices of 2022; variants with segmented network tariffs & Aggregate peak per charging point; diversity factor (import) & Yes. Day-ahead-optimized charging raises the average peak by up to 16\,\% over unoptimized charging & No \\
Flath and Gottwalt \cite{r20} & Germany; microgrid with 900 households, EVs and wind generation; no network model & Other: posted price as a linear function of local net wind; TOU derived from it & Aggregate peak reserve import, 900 households & Yes. Under open-loop prices the reserve peak is about 11 times that under a flat tariff; a closed-loop price, updated after each charging decision, removes it & Yes, in the closed-loop variant \\
Hennig et al. \cite{r36} & Low-voltage neighborhood with 50 German household profiles, up to 25 EVs, 120\,kW transformer & RTP: day-ahead prices of 2020, with fixed, volumetric, capacity subscription or mixed network tariffs & Transformer loading (import) & Yes. Under fixed and volumetric network tariffs the transformer is overloaded from about 8 EVs; a capacity subscription avoids it & No \\
Jones et al. \cite{r2} & US; ten distribution feeders; EV adoption for 2030 & TOU: on-peak period noon to 7\,p.m.; immediate or randomized charging start & Feeder peak (import) & Yes. Immediate charging start raises the residential feeder peak by 20\,\%, new peak at 8\,p.m.; a randomized start lowers it by 5\,\% & No \\
FTI Consulting \cite{r9} & EU; 200 households on one feeder, EV adoption 0 to 60\,\% & RTP: dynamic energy tariff, with four network tariff designs & Aggregate feeder peak (import) & Yes, above 15\,\% EV adoption under a flat volumetric network tariff; capacity-based and dynamic network tariffs cut the peak at 60\,\% adoption by 8 to 24\,\% & No \\
Heinrich et al. \cite{r37} & Denmark (Bornholm); MV feeder, 169 EVs and 150 heaters; transformer rating reduced to 1,200\,kVA & RTP: day-ahead prices, aggregator optimization; capacity limitation service & Primary substation transformer loading (import) & Yes. Night charging at low prices creates off-peak congestion with up to 825\,kW of flexible consumption & No \\
Ziras et al. \cite{r38} & Denmark (DK2); 200 residential 11\,kW EV chargers, charging data 2022 to 2023 & RTP: spot price plus time-of-use network tariff, aggregator cost minimization & Aggregate load of 200 chargers (import) & Yes. Load concentrates between midnight and 3\,a.m., up to about 1,100\,kW, against a historical peak of about 250\,kW for 200 households & No \\
Unterluggauer et al. \cite{r39} & Denmark (Frederiksberg); three meshed urban low-voltage grids with 213 to 562 households; 40\,\% of cars electric, 11\,kW home charging; Monte Carlo power flow for a winter weekday & RTP: hourly spot price plus time-of-use network tariff, cost-minimizing EV charging; time-of-use network tariff alone & Transformer and cable overload probability; peak load (import) & Partly. Charging synchronizes at night, but the peak exceeds that of uncontrolled charging in at most 1.4\,\% (time-of-use) and 0.5\,\% (spot price) of runs. If charging is optimized over several days and forced into the same night, new peaks occur in all runs, with transformer loading above 150\,\% & No \\
Li et al. \cite{r40} & Denmark; RBTS bus 4 test distribution system with 4,779 customers; EV aggregators with Danish driving data; EV penetration 100 and 500\,\% of the household peak & RTP: system locational marginal price; distribution locational marginal prices as alternative & Line loading of the medium-voltage feeder (import) & Yes. Under the system price alone, aggregators charge all EVs in the lowest-price hour and congest the feeder; distribution locational prices spread charging over several hours & No \\
Hennig et al. \cite{r13} & Netherlands; neighborhood of 50 households with 25 EVs; two weeks of January 2021, hourly & RTP: EPEX day-ahead prices, supplier-optimized charging; capacity limitation contracts and capacity subscription as alternatives & Aggregate load of the neighborhood (import) & Yes, qualitatively. Price-optimized charging produces the highest load spikes in low-price hours; capacity limitations spread the load & No \\
Wu et al. \cite{r8} & South Australia; urban feeder with 56 distribution transformers and 2,059 residential customers; EV adoption 50 and 100\,\%; one year of smart meter data & TOU: off-peak from 1\,a.m., adopted by 50 or 100\,\% of EV owners; staggered off-peak start times as alternative & Transformer utilization; overloaded transformers (import) & Yes. Charging at the start of the off-peak window creates an early-morning peak more than 3.5 times the EV peak under a flat tariff; at 100\,\% EVs and full TOU adoption 34 of 56 transformers overload, against 5 under the flat tariff; staggered start times avoid overloads & No \\
Hein et al. \cite{r41} & Germany; SimBench rural low-voltage feeder with 97 buses and a 250\,kVA transformer; EVs (60\,kWh, 11\,kW), heat pumps and PV at 0 to 90\,\% penetration; 15-minute resolution, one week per season & RTP: exogenous 15-minute intraday prices; variants with a time-variable network charge (\S{}14a module 3) and transformer-aware central schedules & Transformer and line loading; voltage (import) & Yes. Price-optimized scheduling raises transformer loading to 253\,\% of rating at 30\,\% penetration, against 38\,\% without EVs and heat pumps and 112\,\% with random charging start; 478\,\% at 90\,\% EV penetration. The time-variable network charge does not lower winter peaks & No \\
Lundblad et al. \cite{r42} & Sweden; synthetic low-voltage grids, about 105,000 grid cells; home charging of 25 to 125\,\% of the vehicle fleet & RTP: hourly spot prices, alone or with a monthly peak power network tariff; logged charging as reference & Peak power; transformer and cable overloads; voltage violations (import) & Yes. At full fleet electrification, average low-voltage peak power rises by 73\,\% with spot-price charging, 51\,\% with logged charging and 10\,\% with the peak power tariff, against no EVs; with spot-price charging, 60\,\% of grid cells have violations from 75\,\% electrification & No \\
Jung and Sundström \cite{r43} & Sweden; real low-voltage grid with 30 customers (Linköping); every household with 10\,kWp PV and an EV, share of EVs at home in daytime 0 to 100\,\%, charging power 3 to 22\,kW, mostly vehicle-to-grid; Monte Carlo simulation & RTP: spot prices of October 2020, with and without a monthly peak-power network charge on the import peak or the absolute peak & Voltage variation (import and export) & Yes. Voltage variation rises markedly with charging power under spot-based tariffs; synchronized selling by many households moves the voltage peak to other hours instead of flattening it & No \\
Wu et al. \cite{r44} & South Australia; 11\,kV feeder with 56 distribution transformers, 2,059 residential and 60 commercial customers; year-round smart meter data; EV penetration up to 80\,\%; TOU adoption 0 to 100\,\%; mixed charging power and frequency & TOU: existing retail time-of-use tariff, charging timed to the start of the off-peak period (around 1 a.m.), against a flat tariff with charging on arrival & Maximum transformer utilization; unserved energy (import) & Yes. Timers synchronize charging at the start of the off-peak period: the peak of a 1,000-EV fleet is 3,688\,kW against 871\,kW under the flat tariff. At 80\,\% EVs and 25\,\% TOU adoption, peak-day transformer utilization falls from 133 to 125\,\%, but on average and high-export days the off-peak spike raises maximum utilization & No \\
Shrestha et al. \cite{r45} & Maryland, US; seven representative feeders clustered from 1,500 feeders of one utility; smart charging enrollment 0 to 50\,\% by 2035 & TOU: charging paused from 5 to 9 p.m.; load balancing as alternative & Feeder peak; transformer overloads; upgrade cost (import) & Yes. At 50\,\% enrollment, TOU cuts feeder peaks by 0 to 12\,\%; at high enrollment the peak moves to the end of the TOU window and forms a second peak, on one feeder raising transformer upgrade cost & No \\
Elmallah et al. \cite{r46} & Northern California, US; 3,043 circuits and 724 substations of one utility; 3.1 million EVs in 2030 rising to 12.5 million in 2050, plus electrified heating & TOU: rate dropping at midnight (base case); spread nighttime charging and daytime charging as alternatives & Feeder and substation capacity exceedance; upgrade cost (import) & Yes. The midnight price drop creates a midnight charging spike; 95 to 260 feeder upgrade projects per year until 2030, against 53 to 87 planned; cumulative upgrade cost 1 to more than 10 billion USD by 2050 & No \\
Unterluggauer et al. \cite{r47} & Denmark (Frederiksberg); urban 10\,kV network with 284 transformers and 440 cables; full electrification of the 2020 car fleet (24,252 resident and 34,818 visiting EVs), 11\,kW home and workplace charging; 40 peak-load weekdays & Other: synchronized start signal: home charging delayed to midnight (stylized low-price period) or 6 p.m., workplace charging to 10 a.m.; against uncontrolled charging & Aggregate peak load; transformers and cables above 100\,\% loading; reinforcement cost (import) & Partly. With non-daily charging, synchronizing at midnight raises the aggregate peak by less than 32\,\%, like uncontrolled charging. If all vehicles charge daily or on the same day, the peak rises by 183 to 215\,\% and 108 to 178 of 284 transformers overload & No \\
Brockmann et al. \cite{r48} & Norway (bidding zone NO5); agent-based model of 200 EVs calibrated on about 28,000 home charging sessions; year 2023 & RTP: day-ahead price plus time-of-use network charge, cost-minimizing charging in each session, with price-sensitive day-to-day plug-in decisions; TOU and flat variants & Coincidence factor of EV fleet charging for 10, 100 and 1,000 EVs, no household load (import) & Yes. Price-sensitive users plug in on the cheapest nights: for 100 EVs the coincidence factor rises from 31\,\% (no price sensitivity) to 46\,\% (high) and 54\,\% (maximum), against 20\,\% under a flat price; the diversity benefit of larger fleets almost disappears & No \\
Brinkel and van Montfoort \cite{r49} & Netherlands; urban grid with 650 households and public charging at 380 stations (196,859 sessions of 2022), scaled to 0 to 100\,\% EV adoption; charge point operators optimize charging & RTP: Dutch day-ahead prices of 2022, with six network tariff designs; fixed retail price as alternative & Distribution grid expansion cost, driven by the aggregate 15-minute peak (import) & Yes. Under day-ahead prices charging concentrates in the cheapest hours with peaks above uncontrolled charging; at 100\,\% adoption grid costs exceed the no-EV level by more than 80\,\%, against 60\,\% under uncontrolled charging & No \\
Winzer and Hensler-Ludwig \cite{r50} & Switzerland (Aargau); 300 households behind one transformer, synthetic profiles; EVs (11\,kW), heat pumps, PV and PV-charged home batteries; device scenarios 2020 (4\,\% of households with an EV) and 2050 (78\,\%) & TOU: network tariff with daily high- and low-tariff periods; CPP; Other: tariff proportional to forecast grid load; RTP: spot-proportional tariff & Aggregate peak, 300 households (import) & Yes. Time-of-use and critical peak tariffs raise the peak above the flat-tariff peak by 5 to 15\,\% in 2020 and by more than 40\,\% in 2050; the grid-load tariff more than doubles it in 2050 & No \\
Iftikhar et al. \cite{r51} & Pakistan (Lahore); 11\,kV residential feeder with 60 transformers and 301 nodes; synthesized seasonal load; 23 or 46 EVs (5 and 10\,\% of feeder peak load) with 1.9\,kW chargers and V2G & RTP: day-ahead hourly price from a national optimal power flow, with aggregator scheduling subject to transformer limits, against a flat rate and a two-part TOU tariff & Transformer loading at weak nodes (import) & Partly. Hourly pricing keeps winter evening loading below the 80\,\% threshold where flat and two-part tariffs exceed it, but loading at one node rises in late-night low-price hours & No \\
Ademulegun et al. \cite{r52} & United Kingdom (Northern Ireland); headroom analysis of 5,620 distribution transformers (100 to 1,000\,kVA); one and two EVs per household & TOU: tariff with night (1 to 8 a.m.) and midday (12 to 4 p.m.) off-peak windows, against charging in normal household demand hours & Share of distribution transformers needing upgrade (import) & No. Off-peak charging cuts the share of transformers needing upgrade from 67 to 52\,\% (one EV per household) and from 78 to 63\,\% (two EVs). Charging within the off-peak windows is not time-resolved, so synchronization at their start is not examined & No \\
Geske et al. \cite{r53} & Luxembourg; 166 synthetic households on a reference transformer (824\,kW import, 742\,kW export limit); 60\,\% with EV, heat pump, PV, home battery and thermal storage; price-responsive share 0 to 60\,\% of households; 2040, hourly, one year & RTP: hourly wholesale prices of the TYNDP Distributed Energy scenario for 2040, plus a capacity-subscription network tariff & Transformer overload hours (import and export) & Yes. Overload hours rise from zero without price-responsive households to 970 per year (11\,\% of hours) when 60\,\% of households respond, 664 in import and 306 in export direction. The average weekday import peak moves from the evening (222\,kW) to midday (480\,kW) & No \\
Lilienkamp et al. \cite{r54} & Germany; 16,191 representative low-voltage grids (402 NUTS-3 regions, six SimBench grid types, eight transformer sizes) aggregated as virtual storage into a European hourly dispatch model; 15 million EVs in 2030 and 35 million in 2045; home charging only, no V2G & RTP: hourly wholesale prices from the dispatch model (endogenous); benchmark run with the same price series fixed exogenously & Transformer loading (charging shifted to respect transformer capacity); transformer expansion need (import and export) & Yes. Price-optimized charging concentrates in low-price hours and adds up to 40\,GW in single hours nationally; to respect transformer limits, up to 7.8\,GW (2030) and 28.8\,GW (2045) of charging must be shifted locally. With exogenous prices the national EV load peak is 75\,\% higher than with endogenous prices. Vehicle-driven transformer expansion need is 1.8 of 16\,GW in 2030 (rooftop PV dominates) and 27.7 of 51.6\,GW in 2045. & Yes; prices are the duals of a European dispatch model \\
Qu et al. \cite{r55} & Stylized setting with a representative UK winter day, hourly; 1,000 EVs in 25 categories; share of price-responsive EVs 0, 30 and 50\,\%; with and without V2G; no grid model & Other: day-ahead retail price set by a profit-maximizing aggregator against an exogenous wholesale price (2 to 20\,ct/kWh, daily mean equal to the wholesale mean); against uncontrolled charging & Aggregate EV load, 1,000 EVs, no household load (import and export) & Partly. Price-responsive EVs charge in the low-price hours 2 to 5 a.m. With 30\,\% price-responsive EVs the night peak stays below the uncontrolled evening peak of about 960\,kW ($-$25\,\% without, $-$7\,\% with V2G). With 50\,\% it exceeds it (+25\,\% without, +55\,\% with V2G). Values read from figures & No \\
\bottomrule
\end{longtable}}
{\footnotesize\setlength{\tabcolsep}{3pt}
\begin{longtable}{@{}>{\raggedright\arraybackslash}p{0.122\linewidth}>{\raggedright\arraybackslash}p{0.174\linewidth}>{\raggedright\arraybackslash}p{0.130\linewidth}>{\raggedright\arraybackslash}p{0.117\linewidth}>{\raggedright\arraybackslash}p{0.390\linewidth}@{}}
\caption{Empirical studies on price-responsive electric vehicle charging and local grid stress}\label{tab:A2}\\
\toprule
\textbf{Study} & \textbf{Setting} & \textbf{Signal} & \textbf{Grid metric} & \textbf{Synchronized new peak?} \\
\midrule
\endfirsthead
\toprule
\textbf{Study} & \textbf{Setting} & \textbf{Signal} & \textbf{Grid metric} & \textbf{Synchronized new peak?} \\
\midrule
\endhead
Van Montfoort et al. \cite{r56} & Netherlands; randomized trial with 600 households with EV chargers & Other: network charge on consumption above 5\,kW, static or TOU, 5\,p.m. to 2\,a.m. & Household peak (withdrawal); aggregate peak, 10 households & Yes, under the time-of-use version. Charging moves to after 2\,a.m. without lowering own peaks; on clusters the new violations outweigh the relief during the priced hours. The static charge has no cluster effect \\
Brylle et al. \cite{r57} & Denmark; metered hourly load of one residential secondary substation with a high EV share (one week, April 2025) and of industrial sites in the Cerius-Radius grid, 2023--2025 & RTP: Nord Pool spot price, plus a static time-of-use network charge & Transformer loading; exceedance events (import) & Yes. 76\,\% of all capacity exceedance events fall in the low-load night window, which is also the low-price window, with load up to 50\,\% above the transformer rating \\
Bailey et al. \cite{r58} & Alberta, Canada; randomized field experiment with 202 EV owners, July to December 2023; EVs grouped on virtual transformers of 10 vehicles with capacities of 12 to 24\,kW & TOU: reward of 3.5\,ct/kWh for home charging from 10\,p.m. to 6\,a.m. and 10\,a.m. to 2\,p.m.; managed charging as second treatment & Transformer capacity violations, virtual transformer with 10 EVs (import) & Yes. Off-peak violations rise by 133\,\% and total violations more than double, exceeding the pre-treatment peak; managed charging halves them \\
Becker et al. \cite{r59} & Flanders, Belgium; 42,089 residential smart meter profiles (862 heat pumps, 1,924 EVs), statistically sampled feeders of 10 to 250 connections; 2022 & TOU: day/night tariff & Sampled feeder peak (import and export); timing of the peak & Yes. Feeders with high-power EV charging peak between 10\,p.m. and 1\,a.m. in winter, which the authors attribute to charging under the night tariff \\
Samende et al. \cite{r60} & UK; charging data of 96 EV households in a living lab, managed by an aggregator; simulated semi-urban (120 households) and urban (205 households) networks with 100 to 500\,kVA transformers; EV penetration 10 to 100\,\% & TOU: lowest price from 11:30 p.m. to 5:30 a.m.; aggregator-optimized charging & Transformer peak (import); voltage & Yes. From about 20\,\% penetration a new peak forms around 1 a.m.; above 30\,\% it exceeds the evening peak, by a factor of up to 2.4 (semi-urban) and 4.0 (urban) \\
Brinkel et al. \cite{r61} & Utrecht, Netherlands; field experiment with public charging stations in five low-voltage grids & RTP: day-ahead prices, optimized by the charge point operator, plus a three-tier network charge by transformer loading & Transformer congestion time (import) & No. Congestion time falls by 21\,\%, from 1.6 to 1.3\,\% of the time, against uncontrolled charging; the low-price tiers are capped to avoid a delayed peak \\
\bottomrule
\end{longtable}}
\twocolumn
\section{Model description}\label{sec:A2}
\begin{table}[!htbp]
\caption{Symbols}\label{tab:A3}
\centering\scriptsize\renewcommand{\arraystretch}{0.92}\setlength{\tabcolsep}{2pt}
\begin{tabular}{@{}>{\raggedright\arraybackslash}p{0.160\linewidth}>{\raggedright\arraybackslash}p{0.500\linewidth}>{\raggedright\arraybackslash}p{0.270\linewidth}@{}}
\toprule
\textbf{Symbol} & \textbf{Definition} & \textbf{Unit or value} \\
\midrule
$t$ & Hour of the month & h \\
$i$ & Household with a price-optimized vehicle & -- \\
$H$ & Number of households with a price-optimized vehicle per grid & 0 to 50 \\
$p_{t}$ & Wholesale price & EUR/MWh \\
$\varphi$ & Network charges and taxes on withdrawal & 10\,ct/kWh \\
$\kappa,\ \kappa^{u}$ & Cycling cost of vehicle batteries, per unit discharged, and of utility-scale storage, per direction & 20\,EUR/MWh each \\
$\eta$ & One-way efficiency & $\surd$0.9 \\
$\overline{P},\ \overline{E}$ & Power and capacity of a vehicle battery & 11\,kW, 60\,kWh \\
${\overline{P}}^{u},\ {\overline{E}}^{u}$ & Power and capacity of utility-scale storage & 10\,GW, 60\,GWh \\
$l_{i,t},\ v_{i,t}$ & Household load and PV generation & kW \\
$w_{i,t},\ e_{i,t}$ & Withdrawal and injection, excluding the refund account & kW \\
$c_{i,t},\ d_{i,t}$ & Charging and discharging, own account & kW \\
$c_{i,t}^{r},\ d_{i,t}^{r}$ & Charging and discharging, refund account & kW \\
$c_{i,t}^{EV},\ d_{i,t}^{EV}$ & Charging and discharging of a vehicle battery, both accounts (A9) & kW \\
$s_{i,t},\ s_{i,t}^{r}$ & State of charge, own and refund account & kWh \\
$m_{i,t}$ & Driving consumption of the vehicle & kWh \\
$h_{i,t}$ & Share of the hour the vehicle is at home & 0 to 1 \\
$\sigma_{i,t}$ & Minimum state of charge of the own account: 80\,\% of capacity at the end of the hour before a departure, 20\,\% while at home, otherwise zero; capped at the level reachable at full charging power & kWh \\
$c_{t}^{u},\ d_{t}^{u},\ s_{t}^{u}$ & Charging, discharging and state of charge of utility-scale storage & GW, GWh \\
$L_{t},\ W_{t},\ S_{t}$ & National load, wind and solar generation & GW \\
${\overline{F}}_{t}$ & Daily mean of net exports and net pumped-storage consumption & GW \\
$E_{t}$ & Storage and vehicle dispatch: discharging minus charging of utility-scale storage and of the vehicle batteries in all N grids & GW \\
$R_{t}$ & Residual load after flexibility (including EVs) & GW \\
$f$ & Price function & EUR/MWh \\
$a$ & Parameter of the cubic price function (4) & 1.8$\times$10$^{-3}$\,EUR/MWh per GW$^{3}$ \\
$a_{2}$ & Parameter of the quadratic price function (5) & 8.9$\times$10$^{-2}$\,EUR/MWh per GW$^{2}$ \\
$\varepsilon_{t}$ & Price residual: observed price minus price function at thermal residual load, fixed across penetrations & EUR/MWh \\
$N$ & Number of identical low-voltage grids & 190,000 \\
\bottomrule
\end{tabular}
\end{table}
This section states the optimization problems of households and utility-scale storage and the equilibrium condition. Table~\ref{tab:A3} defines the symbols. All problems are solved for each calendar month separately.

Each household $i$ with a vehicle minimizes its electricity bill:

\begin{equation}
\begin{split}\min\ \sum_{t}\Bigl[ &\left( p_{t} + \varphi \right)w_{i,t} - p_{t}e_{i,t} + \left( p_{t} + \varphi \right)\left( c_{i,t}^{r} - d_{i,t}^{r} \right)\\ &+ \kappa\left( d_{i,t} + d_{i,t}^{r} \right) \Bigr]\end{split}
\tag{A1}\label{eq:A1}
\end{equation}
Withdrawal and injection balance load, PV generation and the own account:

\begin{equation}
w_{i,t} - e_{i,t} = l_{i,t} - v_{i,t} + c_{i,t} - d_{i,t}
\tag{A2}\label{eq:A2}
\end{equation}
The own account holds solar or grid electricity for driving and for consumption in the household. Driving consumption is deducted from the own account; trips that the battery cannot cover even at full charging power are charged outside the household. The refund account holds grid electricity for re-injection:

\begin{equation}
s_{i,t} = s_{i,t - 1} + \eta c_{i,t} - \frac{d_{i,t}}{\eta} - m_{i,t}
\tag{A3}\label{eq:A3}
\end{equation}
\begin{equation}
s_{i,t}^{r} = s_{i,t - 1}^{r} + \eta c_{i,t}^{r} - \frac{d_{i,t}^{r}}{\eta}
\tag{A4}\label{eq:A4}
\end{equation}
Both accounts share the power and energy limits of the vehicle battery. Charging and discharging are possible only while the vehicle is at home; within an hour, they share the time at home:

\begin{equation}
c_{i,t} + d_{i,t} + c_{i,t}^{r} + d_{i,t}^{r} \leq h_{i,t}\overline{P}
\tag{A5}\label{eq:A5}
\end{equation}
\begin{equation}
s_{i,t} + s_{i,t}^{r} \leq \overline{E}
\tag{A6}\label{eq:A6}
\end{equation}
The own account must hold at least 80\,\% of the battery capacity at the end of the hour before each departure and at least 20\,\% while the vehicle is at home. Both limits are capped at the level that can be reached at full charging power:

\begin{equation}
s_{i,t} \geq \sigma_{i,t}
\tag{A7}\label{eq:A7}
\end{equation}
All variables are non-negative. Each month starts with the own account at 80\,\% of capacity and an empty refund account. Where reachable, the own account ends the month at 80\,\% or more:

\begin{equation}
s_{i,0} = 0.8\overline{E},\ \ s_{i,0}^{r} = 0
\tag{A8}\label{eq:A8}
\end{equation}
Vehicle battery dispatch in Eq. (3) is the sum over both accounts:

\begin{equation}
d_{i,t}^{EV} = d_{i,t} + d_{i,t}^{r},\ \ \ c_{i,t}^{EV} = c_{i,t} + c_{i,t}^{r}
\tag{A9}\label{eq:A9}
\end{equation}
Utility-scale storage maximizes its arbitrage revenue:

\begin{equation}
\max\ \sum_{t}^{}\left[ p_{t}\left( d_{t}^{u} - c_{t}^{u} \right) - \kappa^{u}\left( c_{t}^{u} + d_{t}^{u} \right) \right]
\tag{A10}\label{eq:A10}
\end{equation}
subject to storage dynamics and limits:

\begin{equation}
s_{t}^{u} = s_{t - 1}^{u} + \eta c_{t}^{u} - \frac{d_{t}^{u}}{\eta},\ \ \ c_{t}^{u} + d_{t}^{u} \leq {\overline{P}}^{u},\ \ \ s_{t}^{u} \leq {\overline{E}}^{u}
\tag{A11}\label{eq:A11}
\end{equation}
Under exogenous prices, $p_{t}$ is a fixed time series. Under endogenous prices, $p_{t}$ follows Eqs. (1) to (4) and all agents take it as given. The equilibrium is computed as one linear program per month that minimizes

\begin{equation}
\begin{split}&\sum_{t}\int_{0}^{R_{t}} f(x)\,dx + \sum_{t}\varepsilon_{t}R_{t}\\ &+ N\sum_{i}\sum_{t}\Bigl[ \varphi w_{i,t} + \varphi\left( c_{i,t}^{r} - d_{i,t}^{r} \right) + \kappa\left( d_{i,t} + d_{i,t}^{r} \right) \Bigr]\\ &+ \sum_{t}\kappa^{u}\left( c_{t}^{u} + d_{t}^{u} \right)\end{split}
\tag{A12}\label{eq:A12}
\end{equation}
subject to Eqs. (2), (3) and (A2) to (A11). Its optimality conditions coincide with those of problems (A1) and (A10) at $p_{t} = f\left( R_{t} \right) + \varepsilon_{t}$. The integral is approximated piecewise linearly with 256 segments.

\begin{figure*}[!tp]
\centering
\includegraphics[width=\textwidth]{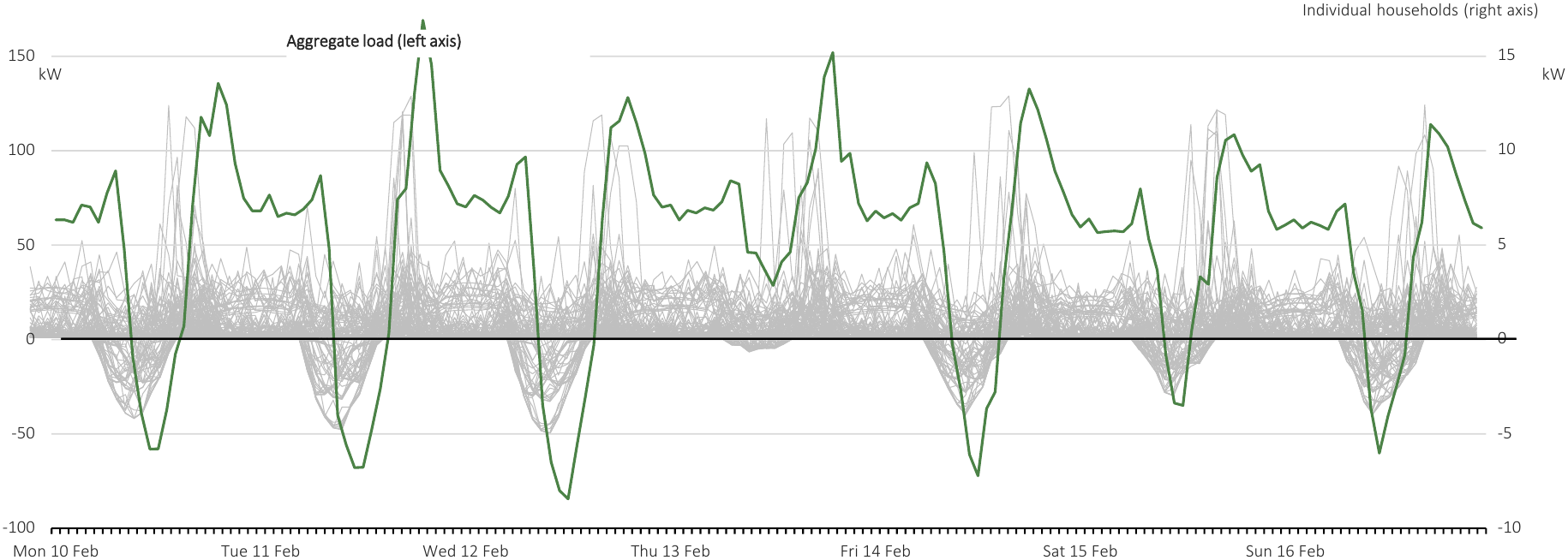}
\caption{Individual household load and aggregate load during the week of the annual import peak (without price-optimized vehicles).}\label{fig:A1}
\end{figure*}
\section{Additional results}\label{sec:A3}
\begin{figure*}[!tp]
\centering
\includegraphics[width=\textwidth]{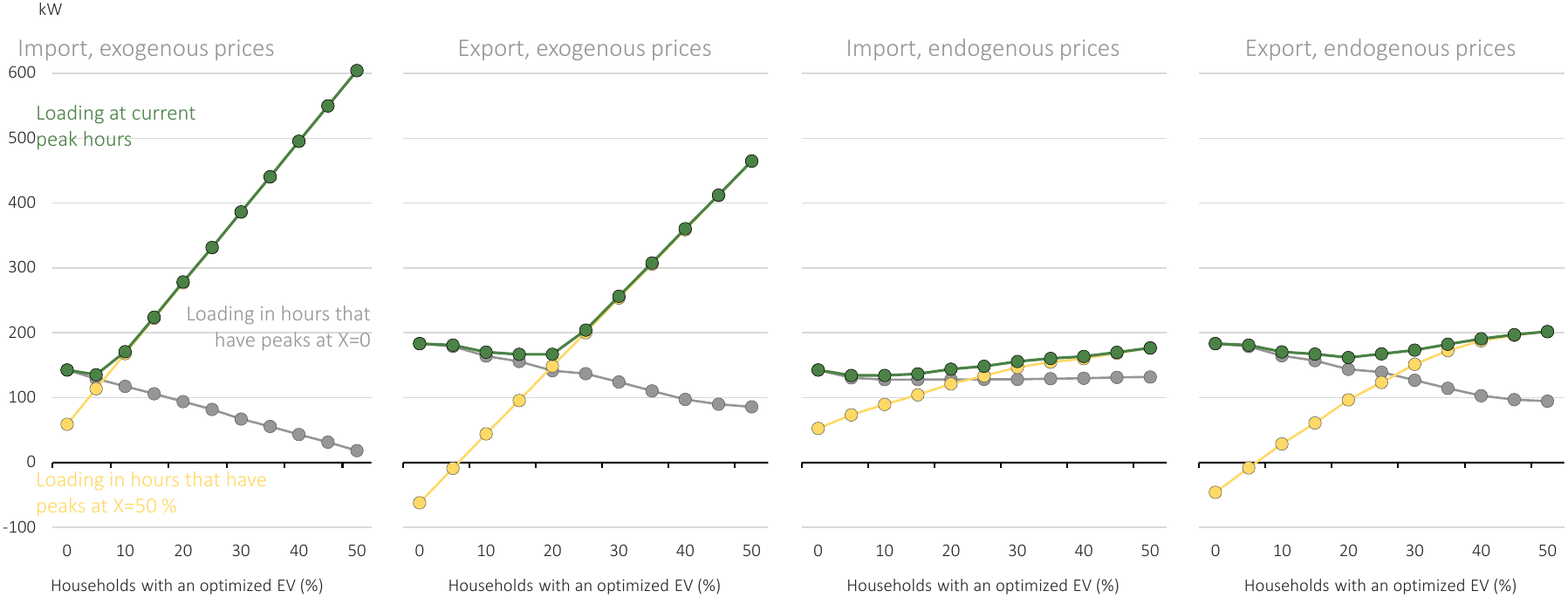}
\caption{Loading in the current peak hours, in the peak hours without optimized vehicles and in the peak hours at full penetration (top 1\,\% of hours).}\label{fig:A2}
\end{figure*}
\begin{figure*}[!tp]
\centering
\includegraphics[width=\textwidth]{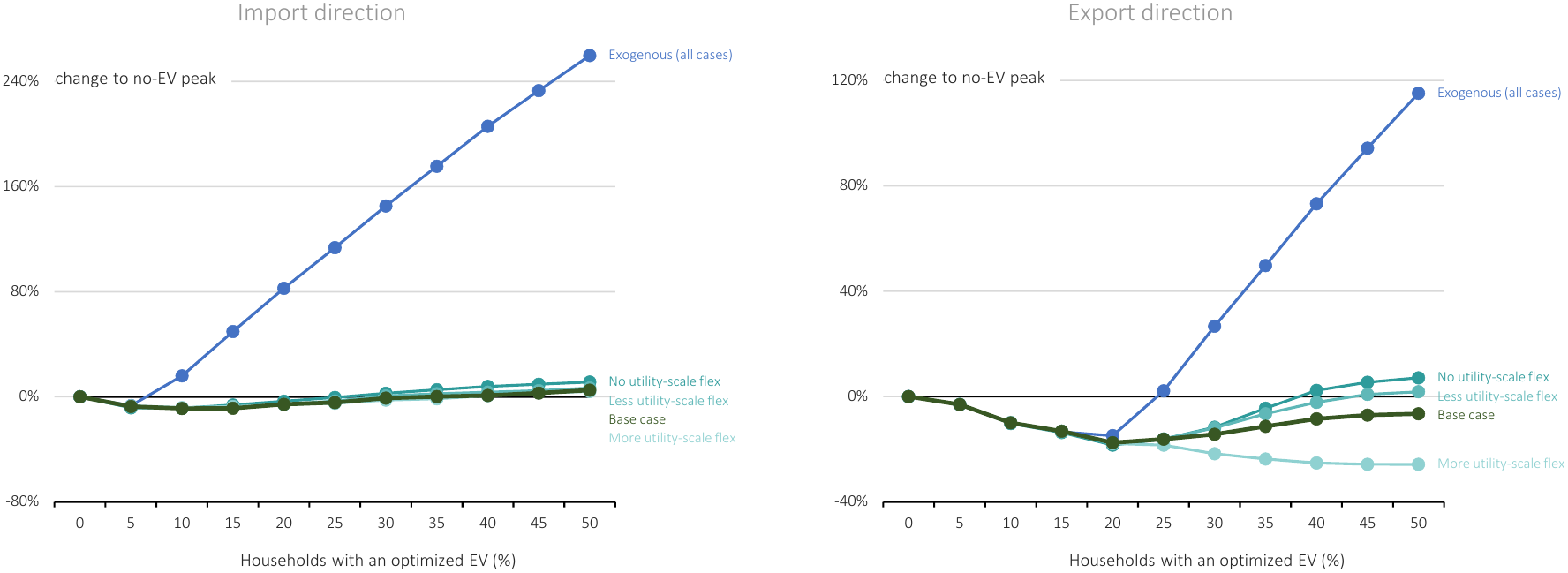}
\caption{Coincident peak for exogenous and endogenous prices: utility-scale flexibility.}\label{fig:A3}
\end{figure*}
\begin{figure*}[!tp]
\centering
\includegraphics[width=\textwidth]{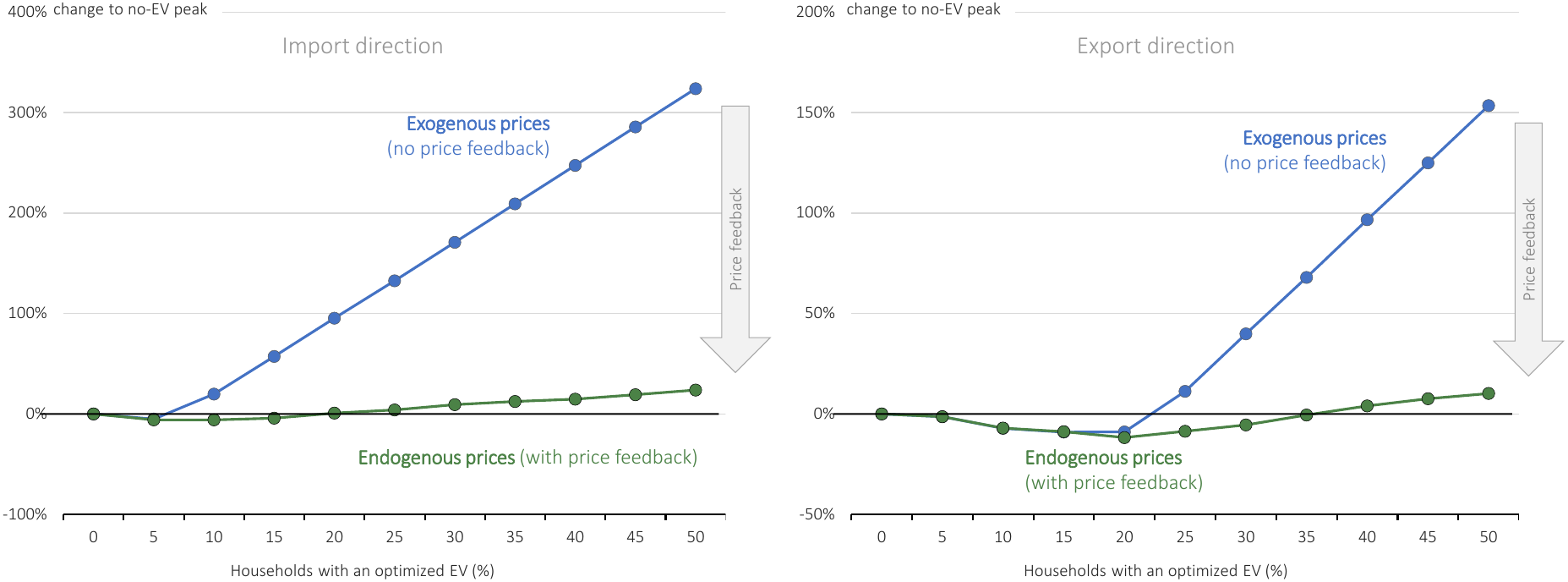}
\caption{Coincident peak for exogenous and endogenous prices, import and export direction, not adjusted for consumption growth.}\label{fig:A5}
\end{figure*}
\section{Electric vehicles and home batteries}\label{sec:A4}
\begin{figure*}[!tp]
\centering
\includegraphics[width=\textwidth]{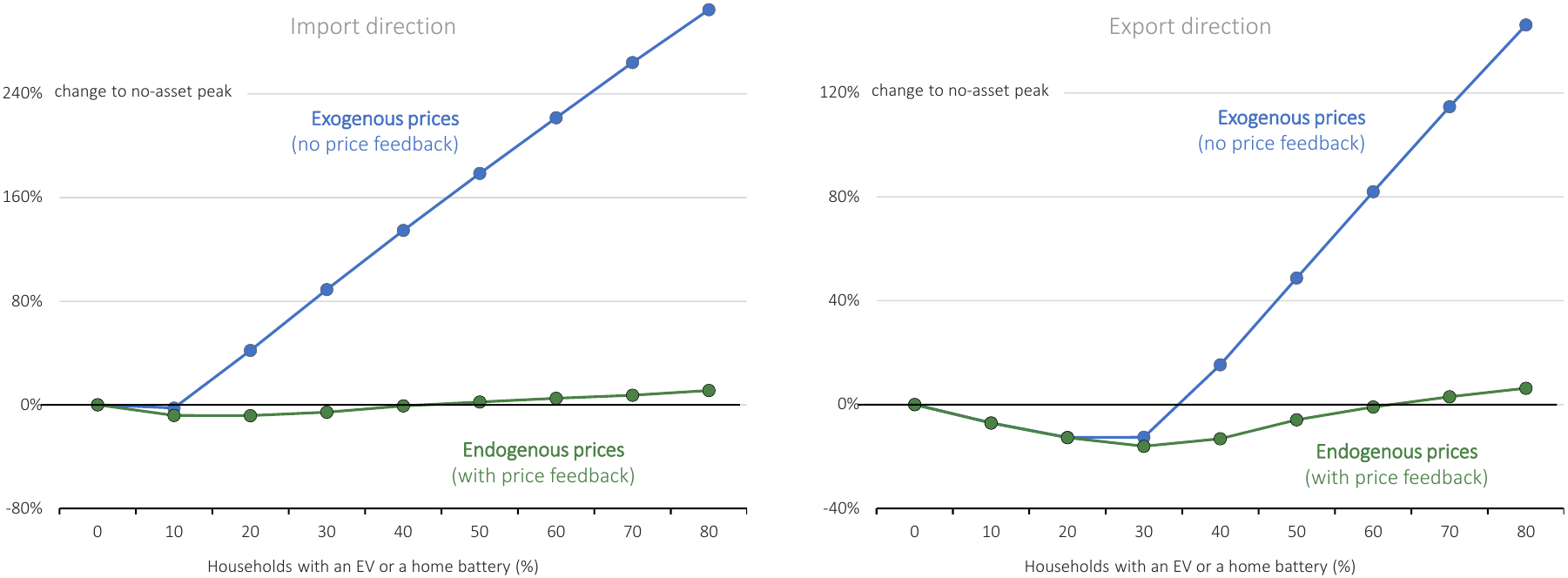}
\caption{Coincident peak for exogenous and endogenous prices: joint roll-out of electric vehicles and home batteries.}\label{fig:A6}
\end{figure*}

Vehicles and home batteries will be adopted side by side. In an additional set of runs, both are rolled out jointly in disjoint households: for a joint penetration of X \% of households, X/2\,\% of households (those with PV) install a home battery of 8\,kWh and 4\,kW, and X/2\,\% of households (those without PV and without an uncontrolled vehicle) get a price-optimized vehicle. X is varied from zero to 80\,\%; both fleets are dispatched in the same equilibrium. Fig.~\ref{fig:A6} and Table~\ref{tab:A5} show the result. Under exogenous prices, the import peak exceeds its initial level from 20\,\% joint penetration and roughly quadruples at 80\,\%; the export peak more than doubles. Under endogenous prices, both peaks first fall; the import peak exceeds its initial level from about 45\,\% and ends slightly above it at 80\,\%, as does the export peak. Without the adjustment for consumption growth, forty vehicles with forty home batteries create about the same import peak as fifty vehicles alone and a higher export peak. The price feedback removes almost the entire increase in both directions.

\begin{table}[!htbp]
\caption{Coincident peak under exogenous and endogenous prices: joint roll-out of electric vehicles and home batteries (change to the peak without vehicles and home batteries, adjusted for consumption growth)}\label{tab:A5}
\centering\scriptsize\setlength{\tabcolsep}{2pt}
\begin{tabular}{@{}>{\raggedright\arraybackslash}p{0.170\linewidth}>{\raggedright\arraybackslash}p{0.310\linewidth}>{\raggedright\arraybackslash}p{0.100\linewidth}>{\raggedright\arraybackslash}p{0.100\linewidth}>{\raggedright\arraybackslash}p{0.100\linewidth}>{\raggedright\arraybackslash}p{0.100\linewidth}@{}}
\toprule
\textbf{Sensitivity} & \textbf{Description} & \multicolumn{2}{>{\raggedright\arraybackslash}p{0.216\linewidth}}{\textbf{Import peak}} & \multicolumn{2}{>{\raggedright\arraybackslash}p{0.216\linewidth}}{\textbf{Export peak}} \\
 &  & \textbf{Exo-}\newline \textbf{genous} & \textbf{Endo-}\newline \textbf{genous} & \textbf{Exo-}\newline \textbf{genous} & \textbf{Endo-}\newline \textbf{genous} \\
\midrule
Joint penetration 20\,\% & 10\,\% of households with a home battery, 10\,\% with a price-optimized vehicle & +42\,\% & -8\,\% & -13\,\% & -13\,\% \\
Joint penetration 40\,\% & 20\,\% of households with a home battery, 20\,\% with a price-optimized vehicle & +135\,\% & -1\,\% & +15\,\% & -13\,\% \\
Joint penetration 60\,\% & 30\,\% of households with a home battery, 30\,\% with a price-optimized vehicle & +222\,\% & +5\,\% & +82\,\% & -1\,\% \\
Joint penetration 80\,\% & 40\,\% of households with a home battery, 40\,\% with a price-optimized vehicle & +305\,\% & +11\,\% & +146\,\% & +6\,\% \\
\bottomrule
\end{tabular}
\end{table}
\begin{table*}[!tp]
\caption{Coincident peak at full penetration under exogenous and endogenous prices, all model runs (change to no-EV peak, adjusted for consumption growth)}\label{tab:A4}
\centering\scriptsize\renewcommand{\arraystretch}{0.95}\setlength{\tabcolsep}{3pt}
\begin{tabular}{@{}>{\raggedright\arraybackslash}p{0.170\linewidth}>{\raggedright\arraybackslash}p{0.385\linewidth}>{\raggedright\arraybackslash}p{0.095\linewidth}>{\raggedright\arraybackslash}p{0.095\linewidth}>{\raggedright\arraybackslash}p{0.095\linewidth}>{\raggedright\arraybackslash}p{0.095\linewidth}@{}}
\toprule
\textbf{Sensitivity} & \textbf{Description} & \multicolumn{2}{>{\raggedright\arraybackslash}p{0.196\linewidth}}{\textbf{Import peak}} & \multicolumn{2}{>{\raggedright\arraybackslash}p{0.196\linewidth}}{\textbf{Export peak}} \\
 &  & \textbf{Exo-}\newline \textbf{genous} \newline \textbf{prices} & \textbf{Endo-}\newline \textbf{genous} \newline \textbf{prices} & \textbf{Exo-}\newline \textbf{genous} \newline \textbf{prices} & \textbf{Endo-}\newline \textbf{genous} \newline \textbf{prices} \\
\midrule
Base case & 50\,\% of households with a price-optimized vehicle; mean load in the highest 1\,\% of hours & +260\,\% & +5\,\% & +115\,\% & -6\,\% \\
\multicolumn{6}{>{\raggedright\arraybackslash}p{0.919\linewidth}}{\textbf{Vehicle and charging parameters} (base case: 60\,kWh / 11\,kW; cycling cost 20\,EUR/MWh)} \\
Large battery & Battery capacity 100\,kWh & +263\,\% & +15\,\% & +120\,\% & +1\,\% \\
Small battery & Battery capacity 40\,kWh & +258\,\% & 0\,\% & +103\,\% & -13\,\% \\
High charging power & Charging and discharging power 22\,kW & +549\,\% & +8\,\% & +308\,\% & -2\,\% \\
Low charging power & Charging and discharging power 5.5\,kW & +102\,\% & 0\,\% & -2\,\% & -24\,\% \\
High cycling costs & Cycling cost of 50\,EUR/MWh on discharge & +174\,\% & -1\,\% & +47\,\% & -20\,\% \\
EVs in all households & Price-optimized vehicles in all 80 households without an uncontrolled vehicle (penetration 80\,\%) & +412\,\% & +14\,\% & +235\,\% & -1\,\% \\
\multicolumn{6}{>{\raggedright\arraybackslash}p{0.919\linewidth}}{\textbf{Charging constraints} (base case: bidirectional charging; 80\,\% state of charge at departure; perfect foresight within each month)} \\
Charging only & No discharging (no vehicle-to-home, no vehicle-to-grid) & +119\,\% & -6\,\% & -19\,\% & -19\,\% \\
Full charge & 100\,\% state of charge at each departure & +216\,\% & -7\,\% & +71\,\% & -22\,\% \\
Daily foresight & One-day optimization horizon instead of one month & +259\,\% & +3\,\% & +111\,\% & -17\,\% \\
Fewer plug-ins & Vehicles plugged in only during home stays that begin on three fixed weekdays per vehicle & +95\,\% & -9\,\% & -15\,\% & -24\,\% \\
V2H only & Discharging only to cover the household's own load; no export from the vehicle & +192\,\% & -2\,\% & -24\,\% & -25\,\% \\
\multicolumn{6}{>{\raggedright\arraybackslash}p{0.919\linewidth}}{\textbf{Local concentration and utility-scale flexibility} (base case: N = 190,000 grids; utility-scale flex 10\,GW / 60\,GWh)} \\
Quarter of the grids & National vehicle fleet 25\,\% of base case & +260\,\% & +78\,\% & +115\,\% & +75\,\% \\
Half of the grids & National vehicle fleet 50\,\% of base case & +260\,\% & +42\,\% & +115\,\% & +53\,\% \\
Larger national fleet & National vehicle fleet 200\,\% of base case & +260\,\% & -12\,\% & +115\,\% & -37\,\% \\
No utility-scale flex & No utility-scale storage & +260\,\% & +11\,\% & +115\,\% & +7\,\% \\
Less utility-scale flex & Utility-scale flexibility halved to 5\,GW / 30\,GWh & +260\,\% & +7\,\% & +115\,\% & +2\,\% \\
More utility-scale flex & Utility-scale flexibility doubled to 20\,GW / 120\,GWh & +260\,\% & +4\,\% & +115\,\% & -26\,\% \\
\multicolumn{6}{>{\raggedright\arraybackslash}p{0.919\linewidth}}{\textbf{Renewables} (base case: 8\,kWp per PV household, no curtailment)} \\
PV curtailment & Economic curtailment of residential PV & +260\,\% & +6\,\% & +115\,\% & -7\,\% \\
Larger residential PV & PV size doubled to 16\,kWp per PV household & +258\,\% & 0\,\% & -4\,\% & -28\,\% \\
More solar & Additional 50\,TWh/a of solar PV (non-residential) & +260\,\% & +13\,\% & +115\,\% & -3\,\% \\
More wind & Additional 50\,TWh/a of onshore wind & +260\,\% & +23\,\% & +115\,\% & -6\,\% \\
Future power system & 2035 scenario: wind 200\,GW, solar 250\,GW, load 750\,TWh, utility-scale flexibility 33\,GW / 200\,GWh & +264\,\% & +43\,\% & +120\,\% & -16\,\% \\
\multicolumn{6}{>{\raggedright\arraybackslash}p{0.919\linewidth}}{\textbf{Network charges} (base case: network charges and taxes 10\,ct/kWh; refund on re-injected electricity)} \\
Low network charge & Network charges and taxes of 5\,ct/kWh & +263\,\% & +10\,\% & +122\,\% & 0\,\% \\
High network charge & Network charges and taxes of 20\,ct/kWh & +252\,\% & +1\,\% & +104\,\% & -12\,\% \\
No charge refund & No refund of the network charge on re-injected electricity & +215\,\% & 0\,\% & +55\,\% & -23\,\% \\
Net metering & Network charge of 10\,ct/kWh only on each household's monthly net withdrawal (withdrawal minus injection); no refund & +266\,\% & +20\,\% & +129\,\% & +7\,\% \\
Injection charge & Network charge of 10\,ct/kWh also on injection; refund on re-injected electricity unchanged & +259\,\% & +4\,\% & +110\,\% & -8\,\% \\
\multicolumn{6}{>{\raggedright\arraybackslash}p{0.919\linewidth}}{\textbf{Price function} (base case: exogenous: historical 2025 prices; endogenous: cubic price function)} \\
Cubic & Cubic price function; exogenous: same price function at 0\,\% EV & +263\,\% & +5\,\% & +115\,\% & -6\,\% \\
Quadratic & Quadratic function; exogenous: same function at 0\,\% EV & +264\,\% & +1\,\% & +113\,\% & -25\,\% \\
Piecewise linear & Piecewise linear function; exogenous: same function at 0\,\% EV & +264\,\% & +1\,\% & +114\,\% & -32\,\% \\
Cubic, $\varepsilon$ = 0 & Cubic function with $\varepsilon$ = 0; exogenous: same function at 0\,\% EV & +261\,\% & -7\,\% & +96\,\% & -35\,\% \\
Flatter price function & Cubic function with $a$ halved; exogenous: same function at 0\,\% EV & +263\,\% & +18\,\% & +114\,\% & +11\,\% \\
Steeper price function & Cubic function with $a$ doubled; exogenous: same function at 0\,\% EV & +266\,\% & -2\,\% & +116\,\% & -26\,\% \\
\multicolumn{6}{>{\raggedright\arraybackslash}p{0.919\linewidth}}{\textbf{Definition of the coincident peak} (base case: mean load in the highest 1\,\% of hours)} \\
Peak hour & Load in the highest hour of the year & +215\,\% & +18\,\% & +111\,\% & +26\,\% \\
Top 10\,\% & Mean load in the highest 10\,\% of hours & +161\,\% & -1\,\% & +51\,\% & -32\,\% \\
\multicolumn{6}{>{\raggedright\arraybackslash}p{0.919\linewidth}}{\textbf{Household simulation} (base case: 100 households, single draw, perfect foresight within each month, hourly)} \\
Bootstrap & 400 replications of the 100 households & +241 $\pm$ 45\,\% & +8 $\pm$ 11\,\% & +115 $\pm$ 25\,\% & -3 $\pm$ 11\,\% \\
Independent networks & 3 grids (new load profiles and coupling), own equilibrium each & +256 to +284\,\% & +4 to +12\,\% & +116 to +117\,\% & -6 to -5\,\% \\
Network size & 25 to 400 households, national fleet unchanged; range & +220 to +281\,\% & +7 to +13\,\% & +107 to +120\,\% & -6 to +1\,\% \\
No load coupling & Household load not coupled to national load & +254\,\% & +4\,\% & +116\,\% & -7\,\% \\
Stronger load coupling & Coupling of household load to national load at double strength & +246\,\% & +2\,\% & +115\,\% & -7\,\% \\
Measured load Zurich & Measured household load (EKZ Zurich, 2024, annual energy normalized) instead of synthetic profiles; against 2024 & +344\,\% & +41\,\% & +139\,\% & +13\,\% \\
Measured load Lucerne & Measured household load (CKW Lucerne) & +334\,\% & +27\,\% & +136\,\% & +4\,\% \\
Quarter-hourly & All quantities quarter-hourly & +248\,\% & +3\,\% & +115\,\% & -7\,\% \\
Historical year & Load, solar, wind and prices of 2023 or 2024 instead of 2025 & +249 to +252\,\% & +6 to +17\,\% & +109 to +133\,\% & -15 to +5\,\% \\
PV orientation & All PV facing south, or half east and half west & +259 to +260\,\% & +4 to +6\,\% & +68 to +156\,\% & -23 to +12\,\% \\
Fewer PV households & 30\,\% of households with PV (base case: 50\,\%) & +260\,\% & +13\,\% & +333\,\% & +85\,\% \\
\bottomrule
\end{tabular}
\end{table*}
\cleanbarrier
\raggedbottom
\section*{Acknowledgment}
I thank Johanna Bronisch, Anselm Eicke and Clemens Lohr for helpful discussions and comments.

\noindent\textbf{Competing interest:} The author is managing director of Neon Neue Energieökonomik, a consultancy advising clients in the electricity sector. This paper did not receive any funding and was not commissioned by any client; no client has seen its results prior to publication.\par\smallskip
\noindent\textbf{Funding:} This research did not receive any specific grant from funding agencies in the public, commercial, or not-for-profit sectors.\par\smallskip
\noindent\textbf{Data availability:} The simulation code, all input data and the complete simulation results are available from the author on request and will be deposited in a public repository upon acceptance.\par\smallskip
\noindent\textbf{CRediT contribution statement:} Lion Hirth: Conceptualization, Methodology, Software, Formal analysis, Investigation, Writing -- original draft, Writing -- review \& editing, Visualization.\par\smallskip
\noindent\textbf{Use of AI:} During the preparation of this work the author used Claude (Anthropic) in order to draft and edit text, write simulation and analysis code, and produce tables and figures. After using this tool, the author reviewed and edited the content as needed and takes full responsibility for the content of the published article.\par

\smallskip

\begin{thebibliography}{99}
\bibitem{r1}
G. Turk, T. Schittekatte, P. Dueñas-Martínez, P.L. Joskow, and R. Schmalensee, ``Designing distribution network tariffs in the US with an application to increased electric vehicle adoption,'' \emph{Energy J.}, vol.~46, no.~6, pp.~143--165, 2025, doi: 10.1177/01956574251365616.

\bibitem{r2}
C.B. Jones \emph{et al.}, ``Impact of electric vehicle customer response to time-of-use rates on distribution power grids,'' \emph{Energy Rep.}, vol.~8, pp.~8225--8235, 2022, doi: 10.1016/j.egyr.2022.06.048.

\bibitem{r3}
E. Veldman and R.A. Verzijlbergh, ``Distribution grid impacts of smart electric vehicle charging from different perspectives,'' \emph{IEEE Trans. Smart Grid}, vol.~6, no.~1, pp.~333--342, 2015, doi: 10.1109/TSG.2014.2355494.

\bibitem{r4}
M. Muratori and G. Rizzoni, ``Residential demand response: Dynamic energy management and time-varying electricity pricing,'' \emph{IEEE Trans. Power Syst.}, vol.~31, no.~2, pp.~1108--1117, 2016, doi: 10.1109/TPWRS.2015.2414880.

\bibitem{r5}
A.D. Hilshey, P.D.H. Hines, P. Rezaei, and J.R. Dowds, ``Estimating the impact of electric vehicle smart charging on distribution transformer aging,'' \emph{IEEE Trans. Smart Grid}, vol.~4, no.~2, pp.~905--913, 2013, doi: 10.1109/TSG.2012.2217385.

\bibitem{r6}
A. Dubey, S. Santoso, M.P. Cloud, and M. Waclawiak, ``Determining time-of-use schedules for electric vehicle loads: A practical perspective,'' \emph{IEEE Power Energy Technol. Syst. J.}, vol.~2, no.~1, pp.~12--20, 2015, doi: 10.1109/JPETS.2015.2405069.

\bibitem{r7}
F. Daneshzand, P.J. Coker, B. Potter, and S.T. Smith, ``EV smart charging: How tariff selection influences grid stress and carbon reduction,'' \emph{Appl. Energy}, vol.~348, Art.~no.~121482, 2023, doi: 10.1016/j.apenergy.2023.121482.

\bibitem{r8}
Y. Wu, S.M. Aziz, M.H. Haque, and T. Kauschke, ``New staggered time-of-use tariffs to mitigate the impact of electric vehicle charging demand on distribution networks,'' \emph{Sustain. Energy, Grids Netw.}, vol.~42, Art.~no.~101682, 2025, doi: 10.1016/j.segan.2025.101682.

\bibitem{r9}
FTI Consulting, ``A roadmap for cost-reflective electricity network tariffs in the EU,'' Report for smartEn, March 2025, 2025.

\bibitem{r10}
S. Backe, G. Kara, and A. Tomasgard, ``Comparing individual and coordinated demand response with dynamic and static power grid tariffs,'' \emph{Energy}, vol.~201, Art.~no.~117619, 2020, doi: 10.1016/j.energy.2020.117619.

\bibitem{r11}
S. Huang, Q. Wu, S.S. Oren, R. Li, and Z. Liu, ``Distribution locational marginal pricing through quadratic programming for congestion management in distribution networks,'' \emph{IEEE Trans. Power Syst.}, vol.~30, no.~4, pp.~2170--2178, 2015, doi: 10.1109/TPWRS.2014.2359977.

\bibitem{r12}
Z. Liu, Q. Wu, S.S. Oren, S. Huang, R. Li, and L. Cheng, ``Distribution locational marginal pricing for optimal electric vehicle charging through chance constrained mixed-integer programming,'' \emph{IEEE Trans. Smart Grid}, vol.~9, no.~2, pp.~644--654, 2018, doi: 10.1109/TSG.2016.2559579.

\bibitem{r13}
R.J. Hennig, L.J. de Vries, and S.H. Tindemans, ``Risk vs. restriction -- An investigation of capacity-limitation based congestion management in electric distribution grids,'' \emph{Energy Policy}, vol.~186, Art.~no.~113976, 2024, doi: 10.1016/j.enpol.2023.113976.

\bibitem{r14}
K. Clement-Nyns, E. Haesen, and J. Driesen, ``The impact of charging plug-in hybrid electric vehicles on a residential distribution grid,'' \emph{IEEE Trans. Power Syst.}, vol.~25, no.~1, pp.~371--380, 2010, doi: 10.1109/TPWRS.2009.2036481.

\bibitem{r15}
E. Sortomme, M.M. Hindi, S.D.J. MacPherson, and S.S. Venkata, ``Coordinated charging of plug-in hybrid electric vehicles to minimize distribution system losses,'' \emph{IEEE Trans. Smart Grid}, vol.~2, no.~1, pp.~198--205, 2011, doi: 10.1109/TSG.2010.2090913.

\bibitem{r16}
P. Richardson, D. Flynn, and A. Keane, ``Optimal charging of electric vehicles in low-voltage distribution systems,'' \emph{IEEE Trans. Power Syst.}, vol.~27, no.~1, pp.~268--279, 2012, doi: 10.1109/TPWRS.2011.2158247.

\bibitem{r17}
J. Hu, S. You, M. Lind, and J. Østergaard, ``Coordinated charging of electric vehicles for congestion prevention in the distribution grid,'' \emph{IEEE Trans. Smart Grid}, vol.~5, no.~2, pp.~703--711, 2014, doi: 10.1109/TSG.2013.2279007.

\bibitem{r18}
J. Quirós-Tortós, L.F. Ochoa, S.W. Alnaser, and T. Butler, ``Control of EV charging points for thermal and voltage management of LV networks,'' \emph{IEEE Trans. Power Syst.}, vol.~31, no.~4, pp.~3028--3039, 2016, doi: 10.1109/TPWRS.2015.2468062.

\bibitem{r19}
A. Dubey and S. Santoso, ``Electric vehicle charging on residential distribution systems: Impacts and mitigations,'' \emph{IEEE Access}, vol.~3, pp.~1871--1893, 2015, doi: 10.1109/ACCESS.2015.2476996.

\bibitem{r20}
C.M. Flath and S. Gottwalt, ``Price-based load coordination revisited: Augmenting open-loop coordination approaches,'' \emph{Bus. Res.}, vol.~9, pp.~157--178, 2016, doi: 10.1007/s40685-016-0033-8.

\bibitem{r21}
L. Gan, U. Topcu, and S.H. Low, ``Optimal decentralized protocol for electric vehicle charging,'' \emph{IEEE Trans. Power Syst.}, vol.~28, no.~2, pp.~940--951, 2013, doi: 10.1109/TPWRS.2012.2210288.

\bibitem{r22}
Z. Ma, D.S. Callaway, and I.A. Hiskens, ``Decentralized charging control of large populations of plug-in electric vehicles,'' \emph{IEEE Trans. Control Syst. Technol.}, vol.~21, no.~1, pp.~67--78, 2013, doi: 10.1109/TCST.2011.2174059.

\bibitem{r23}
M. Roozbehani, M.A. Dahleh, and S.K. Mitter, ``Volatility of power grids under real-time pricing,'' \emph{IEEE Trans. Power Syst.}, vol.~27, no.~4, pp.~1926--1940, 2012, doi: 10.1109/TPWRS.2012.2195037.

\bibitem{r24}
J. Stute and M. Kühnbach, ``Dynamic pricing and the flexible consumer -- Investigating grid and financial implications: A case study for Germany,'' \emph{Energy Strategy Rev.}, vol.~45, Art.~no.~100987, 2023, doi: 10.1016/j.esr.2022.100987.

\bibitem{r25}
L. Semmelmann, K. Kaiser, A. Heider, K.J. Kircher, G. Hug, and C. Weinhardt, ``Analyzing the impact of dynamic tariff adoption and regulatory options on distribution grids with an open-source framework,'' Proceedings of the 16th ACM International Conference on Future and Sustainable Energy Systems (e-Energy 2025), 515--533, 2025, doi: 10.1145/3679240.3734590.

\bibitem{r26}
L. Cong, K. Christensen, M. Værbak, B.N. Jørgensen, and Z.G. Ma, ``Empirically informed multi-agent simulation of distributed energy resource adoption and grid overload dynamics in energy communities,'' \emph{Electronics}, vol.~14, no.~20, Art.~no.~4001, 2025, doi: 10.3390/electronics14204001.

\bibitem{r27}
A. Lilienkamp and N. Namockel, ``Integrating EVs into distribution grids: Examining the effects of various DSO intervention strategies on optimized charging,'' \emph{Appl. Energy}, vol.~378, Art.~no.~124775, 2025, doi: 10.1016/j.apenergy.2024.124775.

\bibitem{r28}
J. Lersch, R. Tang, M. Weibelzahl, J. Weissflog, and Z. Wu, ``Assessing the impacts of energy sharing on low voltage distribution networks: Insights into electrification and electricity pricing in Germany,'' \emph{Appl. Energy}, vol.~378, Art.~no.~124743, 2025, doi: 10.1016/j.apenergy.2024.124743.

\bibitem{r29}
R.A. Verzijlbergh, L.J. de Vries, and Z. Lukszo, ``Renewable energy sources and responsive demand. Do we need congestion management in the distribution grid?'' \emph{IEEE Trans. Power Syst.}, vol.~29, no.~5, pp.~2119--2128, 2014, doi: 10.1109/TPWRS.2014.2300941.

\bibitem{r30}
A. Stawska, N. Romero, M. de Weerdt, and R. Verzijlbergh, ``Demand response: For congestion management or for grid balancing?,'' \emph{Energy Policy}, vol.~148, Art.~no.~111920, 2021, doi: 10.1016/j.enpol.2020.111920.

\bibitem{r31}
C.M. Flath, J.P. Ilg, S. Gottwalt, H. Schmeck, and C. Weinhardt, ``Improving electric vehicle charging coordination through area pricing,'' \emph{Transp. Sci.}, vol.~48, no.~4, pp.~619--634, 2014, doi: 10.1287/trsc.2013.0467.

\bibitem{r32}
P. Huang and M. Sandström, ``Do smart charging and vehicle-to-grid strengthen or strain power grids with rising EV adoption? Insights from a Swedish residential network,'' \emph{Appl. Energy}, vol.~401, Art.~no.~126713, 2025, doi: 10.1016/j.apenergy.2025.126713.

\bibitem{r33}
F. Salah, J.P. Ilg, C.M. Flath, H. Basse, and C. van Dinther, ``Impact of electric vehicles on distribution substations: A Swiss case study,'' \emph{Appl. Energy}, vol.~137, pp.~88--96, 2015, doi: 10.1016/j.apenergy.2014.09.091.

\bibitem{r34}
M. von Bonin, E. Dörre, H. Al-Khzouz, M. Braun, and X. Zhou, ``Impact of dynamic electricity tariff and home PV system incentives on electric vehicle charging behavior: Study on potential grid implications and economic effects for households,'' \emph{Energies}, vol.~15, no.~3, Art.~no.~1079, 2022, doi: 10.3390/en15031079.

\bibitem{r35}
N.K. Panda, N. Li, and S.H. Tindemans, ``Aggregate peak EV charging demand: The influence of segmented network tariffs,'' in \emph{Proc. 2024 IEEE Transp. Electrific. Conf. Expo (ITEC)}, 2024, pp.~1--6, doi: 10.1109/ITEC60657.2024.10599041.

\bibitem{r36}
R.J. Hennig, D. Ribó-Pérez, L.J. de Vries, and S.H. Tindemans, ``What is a good distribution network tariff? Developing indicators for performance assessment,'' \emph{Appl. Energy}, vol.~318, Art.~no.~119186, 2022, doi: 10.1016/j.apenergy.2022.119186.

\bibitem{r37}
C. Heinrich, C. Ziras, T.V. Jensen, H.W. Bindner, and J. Kazempour, ``A local flexibility market mechanism with capacity limitation services,'' \emph{Energy Policy}, vol.~156, Art.~no.~112335, 2021, doi: 10.1016/j.enpol.2021.112335.

\bibitem{r38}
C. Ziras, J.M. Zepter, S. Pechrak, and G. Tsaousoglou, ``Designing a local flexibility market for buying back capacity from electricity consumers connected to the distribution network,'' \emph{Util. Policy}, vol.~93, Art.~no.~101889, 2025, doi: 10.1016/j.jup.2025.101889.

\bibitem{r39}
T. Unterluggauer, F. Hipolito, J. Rich, M. Marinelli, and P.B. Andersen, ``Impact of cost-based smart electric vehicle charging on urban low voltage power distribution networks,'' \emph{Sustain. Energy, Grids Netw.}, vol.~35, Art.~no.~101085, 2023, doi: 10.1016/j.segan.2023.101085.

\bibitem{r40}
R. Li, Q. Wu, and S.S. Oren, ``Distribution locational marginal pricing for optimal electric vehicle charging management,'' \emph{IEEE Trans. Power Syst.}, vol.~29, no.~1, pp.~203--211, 2014, doi: 10.1109/TPWRS.2013.2278952.

\bibitem{r41}
T. Hein, L.P. Wagner, A. Bernhard, and F. Ocker, ``Price-optimized EV and heat-pump scheduling can exceed uncoordinated feeder peaks,'' arXiv:2610.05916, 2026.

\bibitem{r42}
T. Lundblad, M. Taljegård, N. Mattsson, E. Perotti, and F. Johnsson, ``Assessing the impact of vehicle fleet electrification on low-voltage grids: The role of network tariffs in mitigating power system violations,'' SSRN Working Paper 6360789, 2026, doi: 10.2139/ssrn.6360789.

\bibitem{r43}
D. Jung and C. Sundström, ``Analysis of tariffs and the impact on voltage variations in low-voltage grids with smart charging and renewable energy,'' \emph{Energies}, vol.~16, no.~22, Art.~no.~7648, 2023, doi: 10.3390/en16227648.

\bibitem{r44}
Y. Wu, S.M. Aziz, M.H. Haque, and T. Kauschke, ``Comprehensive assessment of electric vehicle charging impact on distribution networks: Integrating diversity in fleet and electricity tariffs,'' \emph{IEEE Access}, vol.~13, pp.~112686--112702, 2025, doi: 10.1109/ACCESS.2025.3584326.

\bibitem{r45}
R.R. Shrestha, Z. Zhou, L. Barua, N. Siddique, K. Balasubramaniam, Y. Zhou, and L. Wang, ``Smart charging impact analysis using clustering methods and real-world distribution feeders,'' 2025 IEEE Power \& Energy Society General Meeting (PESGM), 2025, doi: 10.1109/PESGM52009.2025.11225552.

\bibitem{r46}
S. Elmallah, A.M. Brockway, and D.S. Callaway, ``Can distribution grid infrastructure accommodate residential electrification and electric vehicle adoption in Northern California?,'' \emph{Environ. Res.: Infrastruct. Sustain.}, vol.~2, no.~4, Art.~no.~045005, 2022, doi: 10.1088/2634-4505/ac949c.

\bibitem{r47}
T. Unterluggauer, F. Hipolito, S. Klyapovskiy, and P.B. Andersen, ``Impact of electric vehicle charging synchronization on the urban medium voltage power distribution network of Frederiksberg,'' \emph{World Electric Vehicle Journal}, vol.~13, no.~10, Art.~no.~182, 2022, doi: 10.3390/wevj13100182.

\bibitem{r48}
F. Brockmann, R.S. Kohn, and M. Marinelli, ``When smart charging meets smart users: How price-sensitive plug-in behavior reshapes EV integration,'' \emph{Sustain. Energy Grids Netw.}, vol.~44, Art.~no.~101977, 2025, doi: 10.1016/j.segan.2025.101977.

\bibitem{r49}
N. Brinkel and F. van Montfoort, ``Can grid tariffs be fair and efficient? A comprehensive evaluation of tariff designs for smart electric vehicle integration,'' \emph{Energy Econ.}, vol.~160, Art.~no.~109407, 2026, doi: 10.1016/j.eneco.2026.109407.

\bibitem{r50}
C. Winzer and P. Hensler-Ludwig, ``Design and impact of grid tariffs,'' \emph{Energies}, vol.~17, no.~6, Art.~no.~1364, 2024, doi: 10.3390/en17061364.

\bibitem{r51}
M. Iftikhar, K. Imran, F. Ul Ain, A.K. Janjua, R. Razaq, S. Khan, A. Ilyas, S. Ul Haq, and M. Akmal, ``Smart charging with hourly pricing as policy intervention: Reducing costs and grid congestion for promoting electric vehicles in developing countries,'' \emph{Energy Science \& Engineering}, vol.~14, no.~7, pp.~3472--3486, 2026, doi: 10.1002/ese3.70543.

\bibitem{r52}
O.O. Ademulegun, D. Flynn, B. Oni, M. Bani Mustafa, A.M. Adeyinka, E. Uwimana, and N.J. Hewitt, ``Flexing electric vehicle charging for socio-technical resilience: Distribution transformer headroom and locational value analyses,'' \emph{Sustainable Cities and Society}, vol.~149, Art.~no.~107757, 2026, doi: 10.1016/j.scs.2026.107757.

\bibitem{r53}
J. Geske, B. Ortega Moreno, L. Andolfi, and Z. Yang, ``From duck to turtle: 2040's synchronized demand response reshapes distribution grid load,'' Conference paper, IAEE, Munich, 2026. University of Luxembourg, ORBilu 10993/69208, 2026, \url{https://orbilu.uni.lu/bitstream/10993/69208/1/Geske_et_al_2026_IAEE_Munich.pdf}.

\bibitem{r54}
A. Lilienkamp, N. Namockel, and O. Ruhnau, ``Flexibility in electricity wholesale markets and distribution grids: An integrated model and its application to electric vehicles in Germany,'' EWI Working Paper 25/08, Institute of Energy Economics at the University of Cologne, 2025.

\bibitem{r55}
H. Qu, M. Aldaadi, and G. Verbič, ``Bilevel dynamic pricing for EV aggregators with discrete charging and V2G via deep reinforcement learning,'' \emph{Sustain. Energy Grids Netw.}, vol.~47, Art.~no.~102462, 2026, doi: 10.1016/j.segan.2026.102462.

\bibitem{r56}
F. Van Montfoort, P.T. Dijkstra, M. Mulder, and D. Kopányi, ``Reducing peak load from electric vehicles through grid tariffs: Evidence from a randomized field experiment,'' \emph{J. Environ. Econ. Manage.}, vol.~139, Art.~no.~103354, 2026, doi: 10.1016/j.jeem.2026.103354.

\bibitem{r57}
L. Brylle, N. Andersen, and H. Madsen, ``Scaling demand-side flexibility through dynamic tariffs,'' arXiv:2606.13401, 2026.

\bibitem{r58}
M.R. Bailey, D.P. Brown, E. Myers, B. Shaffer, and F.A. Wolak, ``Electric vehicles and the energy transition: Unintended consequences of time-of-use pricing,'' \emph{American Economic Review: Insights}, vol.~7, no.~4, pp.~550--566, 2025, doi: 10.1257/aeri.20240476.

\bibitem{r59}
T. Becker, R. Smet, B. Macharis, and K. Vanthournout, ``Effect of electric vehicles, heat pumps, and solar panels on low-voltage feeders: Evidence from smart meter profiles,'' \emph{Sustain. Energy, Grids Netw.}, vol.~42, Art.~no.~101705, 2025, doi: 10.1016/j.segan.2025.101705.

\bibitem{r60}
C. Samende, A. Akshaya, S.M. Bhagavathy, J. Melone, R. Snape, and K. Stewart, ``Herding from uncoordinated smart charging of EVs: A real-life demonstration,'' 2025 IEEE PES Innovative Smart Grid Technologies Conference Europe (ISGT Europe), 2025, doi: 10.1109/ISGTEUROPE64741.2025.11305626.

\bibitem{r61}
N. Brinkel, P. Markoti\'{c}, L. Kuiper, S. Warmerdam, B. Baeten, J. Meersmans, H. Fidder, R. Berg, B. van der Ree, W. van Sark, and T. AlSkaif, ``Dynamic grid tariffs for electric vehicle charging: Results from a real-world experiment,'' 2023 IEEE Belgrade PowerTech, 2023, doi: 10.1109/POWERTECH55446.2023.10202869.

\bibitem{r62}
Octopus Energy, ``Intelligent Octopus Go,'' (accessed 25 October 2026), 2026, \url{https://octopus.energy/intelligent-octopus-go}.

\bibitem{r63}
OVO Energy, ``Charge Anytime,'' (accessed 25 October 2026), 2026, \url{https://www.ovoenergy.com/electric-cars/charge-anytime}.

\bibitem{r64}
Jedlix, ``Smart charging platform,'' (accessed 25 October 2026), 2026, \url{https://www.jedlix.com}.

\bibitem{r65}
L. Hirth, ``The missing price feedback: Why studies overstate local grid peaks from synchronized home batteries under real-time pricing,'' arXiv:2609.35111, 2026.

\bibitem{r66}
Statistisches Bundesamt (Destatis), ``43,8 Millionen Wohnungen in Deutschland zum Jahresende 2024,'' Press release No. 336, 17 September 2025, \url{https://www.destatis.de/DE/Presse/Pressemitteilungen/2025/09/PD25_336_31231.html}.
\end{thebibliography}
\end{document}